# Pseudo-noise pulse-compression thermography: a powerful tool for time-domain thermography analysis

Marco Ricci[1*], Rocco Zito, Stefano Laureti

[1]Dipartimento di Ingegneria Informatica, Modellistica, Elettronica e Sistemistica, Università della Calabria,
Via Pietro Bucci, 87036 Arcavacata, Rende CS
e-mail: marco.ricci@unical.it, rocco.zito@dimes.unical.it, stefano.laureti@unical.it

*Abstract*— Pulse-compression is a correlation-based measurement technique successfully used in many NDE applications to increase the SNR in the presence of huge noise, strong signal attenuation or when high excitation levels must be avoided. In thermography, the pulse-compression approach was firstly introduced in 2005 by Mulavesaala and co-workers [1], and then further developed by Mandelis and co-authors that applied to thermography the concept of the thermal-wave radar developed for photothermal measurements [2-3]. Since then, many measurement schemes and applications have been reported in the literature by several groups by using various heating sources, coded excitation signals, and processing algorithms. The variety of such techniques is known as pulse-compression thermography or thermal-wave radar imaging.
Even despite the continuous improvement of these techniques during these years, the advantages of using a correlation-based approach in thermography are still not fully exploited and recognized by the community. This is because up to now the reconstructed thermograms' time sequences after pulse-compression were affected by the so-called sidelobes, i.e. the temperature time trends of the pixels exhibit oscillations, especially in the cooling stage, so that they do not reproduce the output of a standard thermography measurement. This is a severe drawback since it hampers an easy interpretation of the data and their comparison with other thermography techniques.
To overcome this issue and unleash the full potential of the approach, this paper shows how it is possible to implement a pulse-compression thermography procedure capable of suppressing any sidelobe by using a pseudo-noise excitation and a proper processing algorithm.
At the end of the procedure, time-sequences of thermograms are reconstructed that correspond to the sample response to a well-defined virtual excitation, namely a rectangular pulse, making the pulse-compression procedure "transparent". This allows the analysis of pixel time trends by using thermal theory-driven processing such as thermal signal reconstruction, pulsed-phase thermography, etc. Moreover, by tuning the characteristic of the pseudo-noise excitation, it is possible to pass from simulating a very short excitation pulse, retrieving results analogous to pulsed-thermography, to simulating long-pulse excitation to match the sample spectral characteristics maximizing the SNR. This makes the procedure very flexible and extremely attractive in many applications such as high-attenuating materials, characterization of fast thermal phenomena, and inspection of fragile samples inspection, e.g. paintings or other artworks, etc.

I. INTRODUCTION

Infrared thermography (IRT) is a multidisciplinary technique that combines infrared science and heat theory to inspect, monitor and characterize objects and structures by collecting and processing infrared images. IRT is used in nondestructive evaluation (NDE) to assess the quality of industrial samples during production or maintenance, as well as to verify the integrity of complex structures in situ. IRT is also largely utilized in condition monitoring, remote sensing and more and more in medical and cultural heritage diagnostics.

There are two main IRT methods, passive thermography, which is not the subject of the present paper, and active thermography (AT), where a controlled heat source is used to stimulate a thermal response in the sample under inspection (SUT).

The most used AT technique in NDE is the pulsed-thermography (PT), where a short heat pulse (few milliseconds) usually provided by flashlamps generates an almost instantaneous temperature increase on the lighted sample surface [4-5]. The consequent temperature gradient generated within the sample starts a diffusion of the heat from the surface toward the interior. By monitoring the time evolution of the temperature/emissivity of the lighted surface using an IR camera, it is possible to infer information about the inner structure and to detect eventual defects, voids, etc. as any inhomogeneity alters the way heat diffuses, and this influences the temperature of the lighted surface.

From a theoretical point of view, heat diffusion is regulated by the heat equation in 3D. However, for the sake of NDE analysis, if the excitation provides a homogeneous heat flux over all the region-of-interest (ROI), approximating the diffusion as a 1D process guarantees in most cases enough accuracy both in the data processing and in simulations. It is thus generally assumed that the heat flux propagates perpendicularly from the inspected surface, without considering lateral diffusion. Under this assumption, PT is extremely insightful since it allows the impulse responses of each imaged pixel to be reconstructed, and the set of these impulse responses completely characterizes the sample response. In addition, PT is a very simple procedure so, if sufficient sensitivity and SNR are ensured to inspect the sample, PT is the preferable IRT technique among all the possible ones. This fact is further strengthened by the various powerful processing procedures developed for PT data [4-7].

However, PT has limitations in the inspection capability for several reasons: the heat provided with a single flash could be not enough to ensure the SNR requested to detect small and/or deep defects as well as defects characterized by small contrast with respect to the background. At the same time, flash peak power can be too much when inspecting fragile samples, such as thin coating layers or historical paintings, where a too high heat flux concentrated in a short time can irreversibly alter the sample.

For increasing the SNR and hence the detection capability of deep defects, lock-in thermography (LIT) was developed, but at the cost of losing almost the information contained in the impulse responses. Further, the SNR can be optimized only if the optimal inspection frequency is known, and this is not the case in general [8-9]. Step-heating thermography (SHT) is another possible strategy to increase the SNR by delivering more energy to the sample while spreading it in time, but also in this case a lack of resolution in the depth analysis is faced by concentrating the heat at very low frequencies [10-11].

An optimal trade-off between the information amount and the SNR could be achieved by using the so-called long-pulse thermography (LPT) in which the duration of the pulse can be tailored according to the thermal diffusivity of the sample under test (SUT) and the depth at which the thermal contrast must be optimized [12-13].

Pulse durations from several tens of milliseconds to several seconds can be used depending on the case, then tuning the system between PT and SHT and analyzing both the heating and the cooling stages. Flashlamps are not suitable for LPT, whereas LEDs, CW lasers and halogen lamps can be used. This is a positive aspect since on one side such heating systems cannot usually provide very short pulses with enough energy, on the other side, flashlamp systems for PT are quite expensive compared to halogen lamps or LEDs.

Further, for what follows, it is worth noting that LPT is in practice the most common measurement scheme adopted when other types of excitations different from light sources are used, such as eddy currents, vibrations, microwaves, jets, etc. In those cases, the excitation usually delivers a constant heat flux for a certain time, i.e. the time profile of the excitation is a rectangular pulse.

In [14-16] the theoretical analysis of the LPT is reported, together with numerical and experimental comparisons between PT and LPT in terms of contrast, detection capability, and by using different processing methods in the time and frequency domain.

Nonetheless, the considerations about the maximum SNR and the peak power limit done for PT are valid for LPT as well, if short pulses must be used or the sample to inspect is very challenging. Further, given optimal pulse duration, the only means to increase SNR by hardware is to use a larger heat stimulus, with the drawbacks that this can entail in terms of costs, possible nonlinearities, and hazards for fragile samples.

In this scenario, pulse-compression thermography (PuCT), also known in the literature as thermal wave radar imaging (TWRI), was proposed in the last two decades as one of the possible techniques capable of replacing PT in applications where high SNR are requested while preserving the time-spectral information and keeping the heating power under an arbitrary limit [1, 2, 17-22].
However, despite the continuous improvement of these techniques during these years, the advantages of using PuCT have still been not fully exploited and recognized by the community.

This is due to the inherent difficulties in implementing PuCT that up to now made not possible replacing PT with PuCT not only for defects detection but also for their characterization. In most of the PuCT literature indeed, the reconstructed thermograms' time sequences after pulse-compression are strongly affected by the so-called sidelobes, i.e. the temperature time trends of the pixels exhibit oscillations, especially in the cooling stage, so that they do not reproduce the output of a standard PT or a generic AT measurement. This is a severe drawback since it hampers an easy interpretation of the data and their comparison with other thermography techniques. Signal optimization and other tools have been used to mitigate the effect of the sidelobes, but to the best of our knowledge, it is still an open problem to suppress sidelobes in PuCT procedures based on a single measurement.

To overcome this issue and unleash the full potential of the PuCT approach, this paper shows how it is possible to suppress any sidelobe by using a pseudo-noise excitation and a proper processing algorithm. The pivot of this result is that we use PuCT to replace an LPT excitation instead of a PT, as will be explained in the next Sections. In this way, PuCT is "transparent", and its output is not distinguishable from those of an LPT experiment of equivalent SNR.

So, the main novelty of the present paper is that a PuCT procedure based on PN coded excitation has been defined and implemented and, after PuC, the time response of the sample to a well-defined excitation, i.e. a rectangular pulse of arbitrary duration, has been faithfully reconstructed allowing the analysis of time trends based on well-known analytical approaches.

To the authors' knowledge this is the first time such results have been achieved in PuCT. Further, the present PuCT procedure paves the ways for the optimization of many thermography applications, as it will be discussed later.

The paper is organized as follows: the next Section II introduces the generic PuC theory and the specific challenges of PuC while Section III describes the pseudo-noise pulse compression procedure (PN-PuCT) designed and implemented. Section IV reports experimental results obtained in a variety of samples and Section V draws conclusions and perspectives.

## II. THEORY OF PULSE COMPRESSION

The key principle underlying pulse compression (PuC) theory is depicted in Figure 1 and here introduced. Note that it is assumed here and henceforth that the sample under test (SUT), as well as the whole measurement chain, behaves as a linear time-invariant system (LTI), as this is the case in most of the practical applications.

Figure 1-a shows the actual PuC measurement procedure: a coded signal $x(t)$ excites the SUT modelled by its impulse response $h(t)$. Usually, $x(t)$ covers all or almost the bandwidth of the measurement system (sample + measuring apparatus), and its bandwidth is independent by its duration. This is obtained by using frequency-modulated signals or pseudo-noise (PN) binary signals so that the energy delivered is spread in both time and frequency ensuring extreme flexibility in signal excitation design [24-25]. The response $y(t)$ of the SUT to such stimulus is collected and then filtered by the so-called matched filter $\psi(t)$. This last step, which a convolution, represents the pulse compression operation which outputs the signal $\hat{h}(t)$.

Thanks to the commutative property of the convolution, the output of the PuC procedure can be represented as

$$\hat{h}(t) = [\hat{\delta} * h](t) \quad (1)$$

where

$$\hat{\delta}(t) = [x * \psi](t) \quad (2)$$

represents the resolution function of the PuC procedure given the pair $\{x(t), \psi(t)\}$.

Equation 1 can be interpreted in this way: the PuC output $\hat{h}(t)$ is the SUT response to a virtual measurement in which $\hat{\delta}(t)$ is the excitation signal, see Figure 1-b. Since the SNR in the measure of $\hat{h}(t)$ depends on the energy content of $\hat{\delta}(t)$, which in turn is proportional to that of $x(t)$, PuC is a way to measure $\hat{h}(t)$ with high SNR even with limited excitation power as the energy of $x(t)$ can be increased by extending its duration.

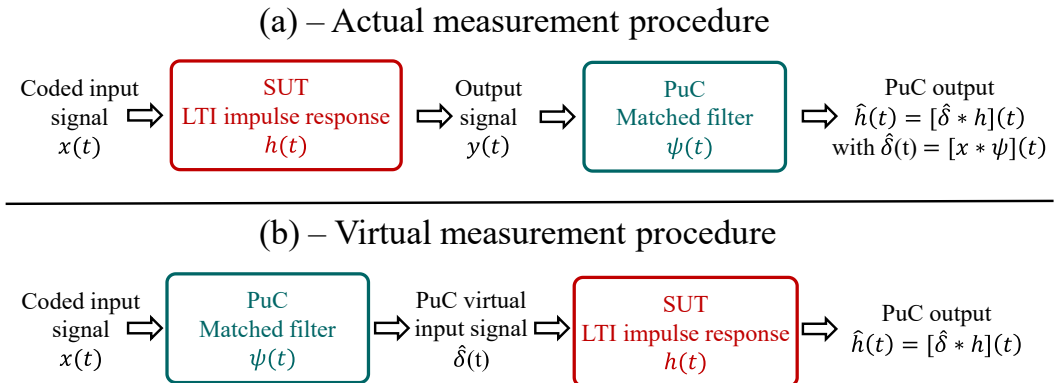

Figure 1: Pulse-compression basic principle. (a) Measurement scheme: a coded signal $x(t)$ excites the sample, which outputs the signal $y(t)$. $y(t)$ is convolved with the matched filter $\psi(t)$ to retrieve the "compressed" signal $\hat{h}(t)$. (b) Equivalent virtual measurement procedure: the same output signal $\hat{h}(t)$ would be obtained by exciting the sample with the excitation signal $\hat{\delta}(t)$, resulting from the convolution between $x(t)$ and $\psi(t)$.

If $\hat{\delta}(t)$ is a Dirac's delta function $\delta(t)$, the PuC output is nothing but the SUT impulse response, $\hat{h}(t) = h(t)$. On the other hand, the more complex is $\hat{\delta}(t)$, the more complex is the analysis of the results.

Unfortunately, $\hat{\delta}(t) \neq \delta(t)$ for any coded signals of finite bandwidth and duration so that, despite the conceptual simplicity of the procedure, the key challenge in any PuC procedure is how to obtain a $\hat{\delta}(t)$ which is optimal for the given application. This requires the contextual optimization of the design of the pair of signals $\{x(t), \psi(t)\}$, and the implementation of the convolution that can be acyclic (single-shot excitation) or cyclic (periodic excitation), see [25].

In general, for NDT methods relying on wave propagation phenomena, and on bandpass measurement chains, a good enough approximation $\hat{\delta}(t)$ of the $\delta(t)$ can be easily obtained as the impulse responses to be characterized are usually sparse, i.e. echograms, and the most important information to get is the location and amplitude of the echoes rather than their shape. This is mostly the case with sonar, radar, ultrasonic NDT, acoustic, and many other techniques.

On the other hand, for NDT techniques relying on diffusion phenomena, such as eddy current and thermography, the impulse responses to be measured are smooth functions, so that even a small level of sidelobes can be detrimental for the faithful reconstruction of $h(t)$ and hence for the successful exploitation of the PuC. In the opinion of the authors, this aspect is very critical in thermography, as the peculiarities of heat phenomena and the characteristics of the common experimental setups, make the implementation of pulse compression thermography (PuCT) not trivial, as detailed below.

As said, for most of the PuC applications including PuCT, two main types of signals are used and sometimes combined: linear and nonlinear frequency modulated "chirp" signals, henceforth referred to as LFM and NLFM respectively, and signals derived by numerical binary codes such as Barker codes (BC), Golay codes (GC), maximum length sequences (MLS), Legendre sequences (LS).

Extensive literature can be found on pulse compression and coded excitations, and a great effort has been devoted to the design $\{x(t), \psi(t)\}$ so that $\hat{\delta}(t)$ guarantees some characteristics. In particular, $\hat{\delta}(t)$ should have in general a main lobe as narrow as possible and sidelobes as small as possible and rapidly decaying [24-26]. So far it was found that if $\psi(t)$ is the time-reversed copy of $x(t)$, $\psi(t) = x(-t)$, the SNR of $\hat{h}(t)$ is maximized [27]. Then, starting from this seminal result, the focus was on the code design optimization to ensure an optimal trade-off between SNR and sidelobes amplitude and energy. For this aim, the matched filter is usually amplitude modulated by a so-called window function $\psi(t) = w(t)x(-t)$. Nevertheless, even with tapering the matched filter or by applying further processing (e.g. Wiener filter [19, 24]), $\hat{\delta}(t)$ exhibits always some sidelobes unless (a) multiple measurements-based strategies, e.g. Golay codes, or (b) periodic PN excitations are used, see [25] for an overview.

In the next Subsection, the main challenges to be faced in order to optimize PuCT are explained.

*II.1 Pulse compression thermography - PuCT*

The full exploitation of PuC benefits in active thermography has not been yet achieved, both due to the inherent difficulties related to the diffusion nature of heat transfer and the correct application of the pulse compression operation, as described in [19].

Two main challenges are well known. The first is the actual unavailability of bipolar heating sources: practical setups use monopolar sources so that coded heating excitations have a DC bias. Again, under the linear hypothesis, the time signals of single pixels are a superposition of the expected response to a bipolar coded excitation (henceforth referred to as AC component) and the response to a step-heating/long-pulse excitation depending on the measurement scheme (henceforth referred to the DC component). The removal of this DC component before the convolution with the matched filter is fundamental for the correct application of the PuC and several approaches have been developed and reported, relying on the use of proper fitting functions [19] or two-measurements-based procedures [28].

The second issue, which is the key one, is the presence and the need to suppress the mathematical noise – i.e. sidelobes – in $\hat{\delta}(t)$ that could arise from the correlation operation.

This is because the sidelobes of $\hat{\delta}(t)$ makes the PuCT output signals apparently not "physically sound" since unexpected oscillations of the reconstructed pixels' temperature trends appear whereas a monotonic behavior is expected. Thus, the processing methods developed for PT and LPT cannot be applied straightforwardly.

In most of the PuCT/TWR applications reported in literature, just a few features are indeed usually extracted from the PuCT output pixels' signals such as the cross-correlation peak value and peak delay time [1-2,17-18], while analysis procedures on the whole time-domain signals are rarely implemented [19-20].

A generic PuCT procedure highlighting these two aspects is sketched in Figure 2, while Figure 3 reports some examples of standard signals used in PuCT together with their autocorrelation functions $\hat{\delta}(t)$: an LFM and an NLF chirp, a Barked-code waveform, and a pair of Golay sequences. From subplots 3.a and 3.b, and by having in mind Eq.2, it is easy to figure out why PuCT output signals are far from those obtained in PT or LPT experiments.

Figure 3.c illustrates an example of a Golay code pair, used in multiple-measurement procedure. A Golay code of a given length $L = 2^K$ consists of a pair of binary sequences having complementary autocorrelation functions such that the main autocorrelation lobe is the same for both while the sidelobes are opposite. By doing two PuC experiments, one for each sequence of the pair as excitation, and summing the two PuC outputs, the sidelobes are virtually suppressed if the system is linear [29-30]. Note also that the sidelobes peak level of the individual Golay sequences is larger than that of the Barker one. This is why in a "singe-sequence single-shot" PuCT scheme Barker codes are preferred among other binary excitations [19].

Golay codes have been applied to PuCT, see for instance [31], but as a matter of fact the sidelobes suppression was not obtained. Once again, this is because PuCT application is not so straightforward. Both the removal of the DC component and the correct mathematical procedure for implementing correlation are fundamental. Further, the application of Golay pairs assumes that the two measurements start exactly from the same initial state, and hence from the same temperature. This requires waiting a certain time between the two measurements, making the procedure less effective, especially if the goal is to detect deep defects that need quite a long excitation to reach the necessary SNR. However, a similar strategy based on a pair of "complementary" signals has been correctly implemented in [28] to remove the DC component, so paving the way for a correct implementation of Golay-based PuCT.

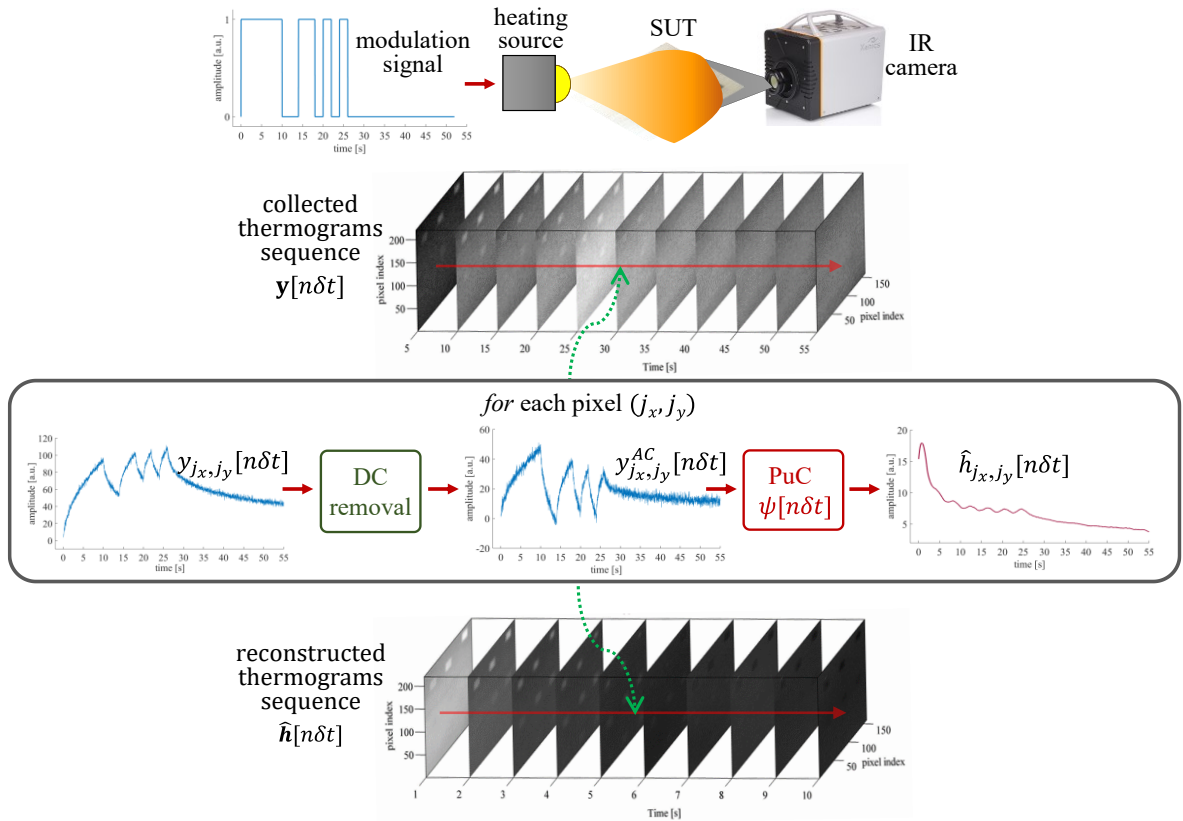

Figure 2: Examples of a PuCT procedure: (1) a coded excitation waveform (13-bit Barker code) modulates the amplitude of a heating source that heats up the sample; (2) an IR camera collects a sequence of thermograms during the heating/cooling stages, denoted by the 3D variable $\mathbf{y}[n\delta t]$. The thermograms are acquired at a fixed rate $FPS = \frac{1}{\delta t}$ (3) for each pixel the PuCT procedure (DC removal + PuC) is implemented on the pixel intensity time-trend $y_{j_x,j_y}[n\delta t]$; (4) a "compressed" thermograms' sequence is reconstructed where $\hat{h}_{j_x,j_y}[n\delta t]$. Note the oscillations in the $\hat{h}_{j_x,j_y}$, which are due to the sidelobes of $\hat{\delta}(t)$, see Figure 3.

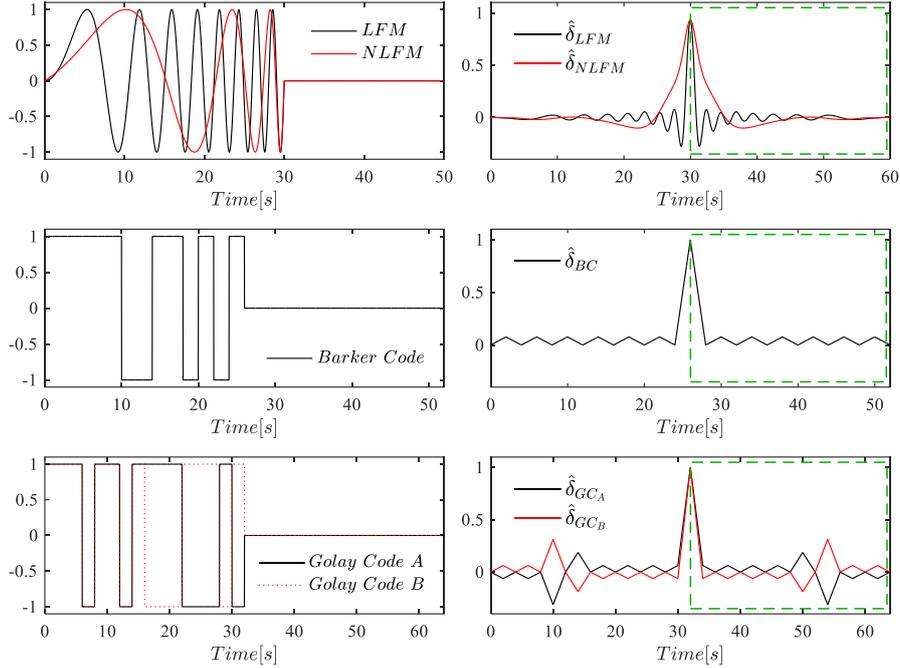

Figure 3: Examples of coded excitations (left column) used in PuCT and the relative resolution functions $\hat{\delta}(t)'s$ (right columns): (a) 30s-long LFM (0-0.5 Hz) and NLFM (0.005-0.5 Hz) defined as in [20] for a thermal diffusivity $\alpha = 10^{-6}\left[\frac{m^2}{s}\right]$; (b)13-bit length Barker code (single bit duration 2s); (c) 16-bit long Golay code pairs (single bit duration 2s). Only the right part of the autocorrelation enclosed in the green dashed rectangle must be considered as resolution function for PuC applications. Note that Barker codes of length > 13 bits do not exist, and BC are the only binary finite length sequences having an autocorrelation with only positive sidelobes and with the lowest possible amplitude, except for the use of two complementary sequences such as Golay codes or of a periodic PN excitation.

In this scenario, we developed a PuCT procedure "sidelobes-free" which exploits the use of PN sequences (MLS and LS) and their cyclic autocorrelation properties.

The procedure is explained in the following Sections, and it represents the extension to thermography of methods previously developed by some of the authors for ultrasound [32] and eddy current [33], so the reader is referred to those articles for an extensive theoretical explanation of the mathematical background and the other features of the PN-based PuC procedure such as the demonstration of how it can overcome and suppress the quantization noise [32].

Nonetheless, for the peculiarities of the PuCT previously mentioned, key changes and tuning of the excitation signals and the data processing were required.

In the next Sections, the PN-PuCT procedure is explained firstly introducing the key idea, then summarizing the main properties of PN sequences, and eventually describing step by step the practical implementation.

It is worth nothing that PN excitation waveforms based on MLS codes were also used in thermography by Bodnar and colleagues to inspect cultural heritage items, as reported in [34-36].

However, in those articles PuC was not exploited and the estimate of the system impulse response was obtained by using an ARMA model and a least-square optimization approach for the system characterization. One calculated the ARMA parameters, some features in both time and frequency domain were calculated.

Nonetheless, also in those cases, an analysis of the whole pixels time signals was not carried out.

III. PSEUDO NOISE PULSE-COMPRESSION THERMOGRAPHY: THEORETICAL ASPECTS

*III.1 PN-PuCT procedure*

The key point of the present PN-PuCT procedure is that after its application we want to retrieve the response of the sample to a rectangular pulse $\Pi(t,T)$ of duration T and arbitrary amplitude A:

$$\text{LPT excitation signal x(t)} = \Pi(t,T) = \big(\theta(t) - \theta(t-T)\big) \quad (3)$$

where $\theta(t)$ is the Heaviside step function.

To achieve this goal, according to Eq. 2, we must find a pair of $x(t), \psi(t)$ such that:

$$\{x * \psi\}(t) = \hat{\delta}(t) = \Pi(t,T) \quad (4)$$

If we can do that, the PuCT procedure is "transparent" with respect to the LPT excitation, but it reduces the noise. Alternatively, PuCT can be seen as a virtual amplifier so that at the end of the procedure the output to an excitation signal $G \times \Pi(t, T)$ is obtained where $G$ is the PuCT gain.

The condition expressed by Eq.4 can be satisfied by exploiting the periodic, i.e. cyclic, correlation properties of PN sequences such as MLS and LS.

Another important aspect to consider is the following: PuC theory is usually introduced for analog signals, as in Eqs. 1-4, and the thermal excitation signal which modulate the heat source is analogue as well, however the PuC algorithm deals with discrete-time signals, namely the pixels intensity along a sequence of thermograms acquired at a fixed frame rate, i.e. $FPS$, and the correlation properties of PN codes here exploited are defined for discrete time signals. Therefore, care must be taken in considering both the $FPS$ of the thermograms acquisition and the sampling frequency $f_{Bit}$ of the PN codes. In the next Subsection the PN sequences used for PuCT are introduced together with their autocorrelation properties, then different PuCT strategies are defined depending on $FPS$ is a multiple or equal to $f_{Bit}$.

### III.2 PN sequences

PN sequences are discrete-time signals, usually binary o ternary, having spectral and randomness properties like those of white noise. There is a great variety of and a huge literature about PN sequences and their many applications ranging from telecommunication [37], to acoustics [38], passing by medical diagnostics [39], and many much more, including NDE.

For the present application, we seek for PN sequences having: (i) an ideal cyclic autocorrelation, and hence a flat spectrum, allowing for the equation (4) to be satisfied and (ii) a variable length to cope with experimental constraints and needs. Among all the possible PN types, modified Maximum Length Sequences and Legendre sequences satisfy such requirements.

### III.2.1 Maximum Length Sequences

Maximum Length Sequences, known as $m$-sequences or $MLS$, are $PN$ binary codes which can be generated as a periodic sequence of "0's" and "1's" by a linear-feedback shift register (LFSR) machine with $M$ taps whose values are derived from the coefficients of a primitive polynomial $p(x)$ of order $M$ in the Galois field $GF(2^M)$. Each $MLS$ is indeed associated with a recursive equation such as:

$$MLS[n] = \left(\sum_{i=1}^{M} \alpha_i MLS[n-i]\right)_{mod(2)} \quad for\ MLS[n] \in \{0,1\}; \tag{5}$$

where $\alpha_i: \{0,1\}$ are the coefficients of $p(x)$:

$$p(x) = x^M + \sum_{i=0}^{M-1} \alpha_i x^i \tag{6}$$

For PuC applications, binary bipolar sequences are of interest, that is $MLS[n] \in \{1, -1\}$ (where "0" is mapped into "1" and "1" into "-1") so Equation 5 becomes:

$$MLS[n] = \prod_{i=1}^{M} MLS[n-i]^{\alpha_i} \quad for\ MLS[n] \in \{1, -1\}; \tag{7}$$

By starting from any nontrivial $M$-tuple of bits, i.e. anyone with not all 1's, a periodic sequence is generated. $MLS$s exhibit the longest possible length for a given $M$-tap shift register equal to $N_{bit} = 2^M - 1$, here the origin of the names. For more details on $MLS$s properties, refer to [32, 40-41]. While the generation by means of LFSR is easy and can be done in real time, finding the primitive polynomial is a hard task, especially for large $M$'s, but fortunately lists of primitive polynomials over $GF(2^M)$ are tabulated in literature up to very high orders, see for instance [42].

For clarity, from now on, we denote with $\{MLS_{N_{bit}}[n]: n \in [0, N_{bit} - 1]\}$ a sequence of length $N_{bit}$ and we will use also an overbar, e.g., $\overline{MLS_{N_{bit}}}[n]$, to indicate a periodic repetition of the single period.

### III.2.2 Legendre Sequences

Differently from $MLS$, Legendre sequences, also known as $\ell$-sequences and henceforth denoted as $LS_{N_{bit}}[n]$, exist for any value of $N_{bit}$ which is a prime number, so many more lengths are available compared to $MLS$ and this is why we prefer for the experimental applications to use $LS$ instead of $MLS$, even if $MLS$ have many other properties that can be exploited.

A $LS_{N_{bit}}[n]$ cannot be generated by a LFSR but it is defined by the Legendre symbol $\left(\frac{n}{N_{bit}}\right)$ as for Equation 8 [38, 42-43]

$$LS_{N_{bit}}[n] = \left(\frac{n}{N_{bit}}\right) \equiv \begin{cases} 0 & \text{if } n \equiv 0 \pmod{N_{bit}} \\ 1 & \text{if } n \text{ is a quadratic residue} \pmod{N_{bit}} \\ -1 & \text{if } n \text{ is not a quadratic residue} \pmod{N_{bit}} \end{cases} \quad (8)$$

$LS$ are "almost" binary as they consist of all "+1" and "-1" except for the first value which is always a "0". Apart from this initial "0", and the different lengths available, MLS's and LS's have very similar autocorrelation and spectral properties, so the PuCT procedure works with both in the same way.

In Table 1 some MLS and LS sequences are reported for different $N_{bit}$ values.

Table 1: Examples of $MLS_{N_{bit}}[n]$ and $LS_{N_{bit}}[n]$ sequences

| $MLS_{N_{bit}}[n]$ | $LS_{N_{bit}}[n]$ |
|---|---|
| $N_{bit} = 7: MLS_{N_{bit}}[n] = \{+,-,-,+,-,+,+\}$ | $N_{bit} = 7: LS_{N_{bit}}[n] = \{0,+,+,-,+,-,-\}$ |
| $N_{bit} = 15: \{+,+,+,-,+,+,-,-,+,-,+,-,-,-,-\}$ | $N_{bit} = 11: \{0,+,-,+,+,+,-,-,-,+,-\}$ |
| $N_{bit} = 31: \{+,+,+,+,-,+,+,-,+,-,-,+,+,-,-,$ $-,-,-,+,+,+,-,-,+,-,-,-,+,-,+,-\}$ | $N_{bit} = 31: \{0,+,+,-,+,+,-,+,+,+,+,-,-,-,+,$ $-,+,-,+,+,+,-,-,-,-,+,-,-,+,-,-\}$ |

*III.2.3 PN sequences with perfect cyclic autocorrelation*

$MLS$ and $LS$ sequences are known for their almost ideal cyclic/periodic autocorrelation function (PACF). The cyclic autocorrelation of an arbitrary finite sequence $s[n]$ is obtained when a periodic repetition of $s[n]$ is convolved with a time-reversed copy of $s[n]$, as illustrated in Figure 4 for $s[n] = MLS_7[n]$. After an initial transient period of length $N_{bit}$, the cyclic autocorrelation is periodic, and the period of the steady-state autocorrelation is the PACF of the $s[n]$ sequence.

The PACF of $s[n]$ is usually denoted as $\Phi_s[n]$ and it can be calculated by exploiting the convolution theorem for discrete time signal as follows [44]:

$$\Phi_s[n] = IDFT\{|DFT\{s[n]\}|^2\} = IDFT\{DFT\{s[n]\}.\times conjugate\{DFT\{s[n]\}\}\} \quad (9)$$

Where, $DFT\{\cdot\}$ is the Discrete Fourier Transform, $IDFT\{\cdot\}$ is the Inverse Discrete Fourier Transform, ".×" stands for pointwise multiplication, $conjugate\{\cdot\}$ returns the complex conjugate of each element.

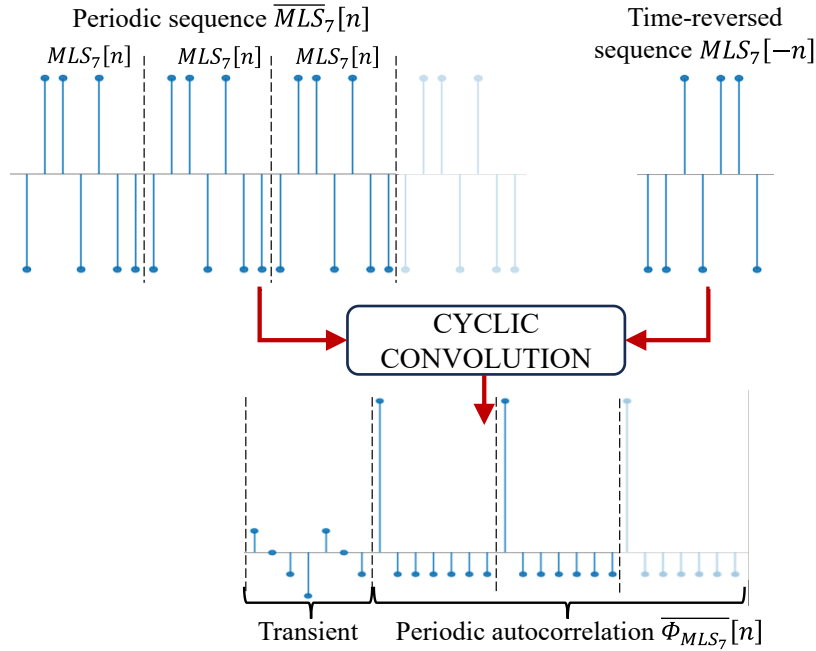

Figure 4: graphical representation of the cyclic convolution between a periodic PN sequence (here an $MLS_7[n]$) and the time-reversed copy of its single period. The convolution output exhibits a transient of $N_{bit}$ and then it become periodic with period equal to $N_{bit}$. A single period of the steady state is the so-called PACF of the PN sequence.

It is worth highlighting that $conjugate\{DFT\{s[n]\}\}$ is nothing but the Discrete Fourier Transform of the time reversed sequence: $conjugate\{DFT\{s[n]\}\} = DFT\{s[-n]\}$, so we retrieve that $\Phi_s[n]$ is given by the cyclic convolution between the periodic sequence $\bar{s}[n]$ and the time-reversed copy of the single period $s[-n]$, henceforth denoted by $\check{s}[n]$:

$$\Phi_s[n] = \{s * \check{s}\}_{cyc}[n] \qquad (10)$$

If we replace $\Phi_s[n]$ with $\hat{\delta}[n]$, Equation 10 is the discrete time counterpart of Equation 2, where $\check{s}[n]$ acts as the matched filter, retrieving the Turin's result.

The PACF of $MLS's$ and $LS$'s approximates a truncated unit pulse signal $\delta_{N_{bit}}[n]$ of length $N_{bit}$ defined as:

$$\delta_{N_{bit}}[n] = \begin{cases} 1 \text{ for } n = 0 \\ 0 \text{ for } n \neq 0 \end{cases}, \ n \in [0, N_{bit} - 1] \qquad (11)$$

Precisely, it is well known that the PACFs for $MLS$ and $LS$ sequences are expressed by:

$$\Phi_{MLS}[n] = \begin{cases} N_{bit} \text{ for } n = 0 \\ -1 \text{ for } n \in [1, N_{bit} - 1] \end{cases} = (N_{bit} + 1)\delta_{N_{bit}}[n] - 1 = (N_{bit} + 1)(\delta_{N_{bit}}[n] - \frac{1}{N_{bit}+1}) ; \qquad (12)$$

$$\Phi_{LS}[n] = \begin{cases} N_{bit} - 1 \text{ for } n = 0 \\ -1 \text{ for } n \in [1, N_{bit} - 1] \end{cases} = N_{bit}\delta_{N_{bit}}[n] - 1 = N_{bit}(\delta_{N_{bit}}[n] - \frac{1}{N_{bit}})$$

As $N_{bit}$ increases, $\Phi_{MLS}[n]$ and $\Phi_{LS}[n]$ become more and more a good approximation of a $\delta_{N_{bit}}[n]$ except for a multiplicative factor, and this property was exploited in the previous PuC procedures based on PN codes [25,32-33].

However, in PuCT applications $N_{bit}$ is not so large and, further, also a very small constant level of sidelobes can be detrimental in the reconstruction of the PuCT output, so we modified the MLS and LS sequences to achieve a perfect PACF. This can be done by adding a proper constant bias to the sequences, or by replacing all "-1" with a different value "q" as reported in [45].

For the PN-PuCT procedure, we modify the $MLS$ and $LS$ sequences as follows to achieve an ideal PACF:

$$\begin{cases} MLS_{N_{bit}}[n] \rightarrow MLS_{N_{bit}}[n] + \dfrac{1 + \sqrt{N_{bit} + 1}}{N_{bit}} = MLS^+_{N_{bit}}[n] \\ LS_{N_{bit}}[n] \rightarrow LS_{N_{bit}}[n] + \dfrac{1}{\sqrt{N_{bit}}} = LS^+_{N_{bit}}[n] \end{cases} \qquad (13)$$

Precisely, for the modified sequences the following expressions hold:

$$\Phi_{MLS^+_{N_{bit}}}[n] = \begin{cases} N_{bit} + 1 \text{ for } n = 0 \\ 0 \text{ for } n \neq 0 \end{cases} = (N_{bit} + 1)\delta_{N_{bit}}[n] ;$$

$$\Phi_{LS^+_{N_{bit}}}[n] = \begin{cases} N_{bit} \text{ for } n = 0 \\ 0 \text{ for } n \neq 0 \end{cases} = N_{bit}\delta_{N_{bit}}[n] \qquad (14)$$

Figure 5 shows an example of $LS$ and $MLS$ sequences for $N_{bit} = 31$ together with the corresponding PACFs; Figure 6 depicts the modified sequences $LS^+$ and $MLS^+$ and their relative perfect PACFs, always for $N_{bit} = 31$

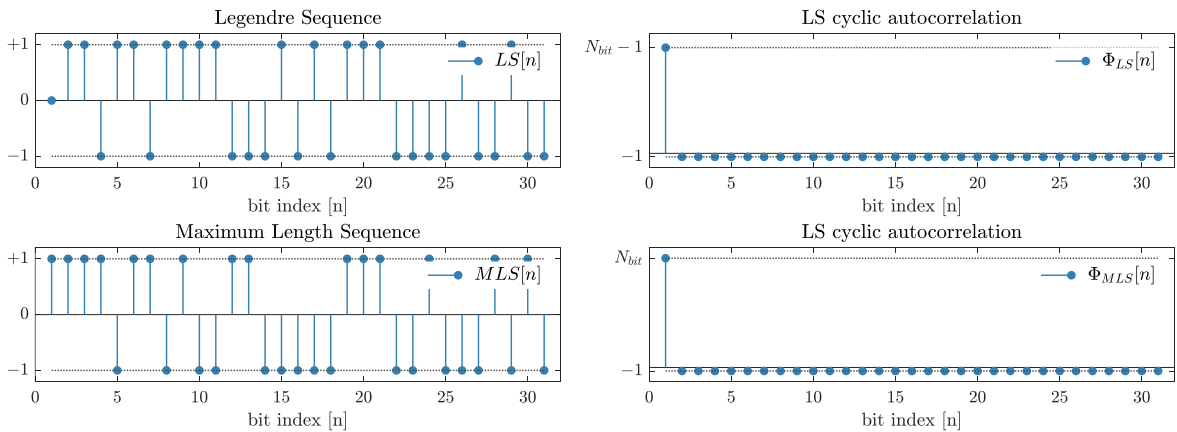

Figure 5: (left) an example of $LS$ and $MLS$ sequences for $N_{bit} = 31$; (right) the corresponding PACFs

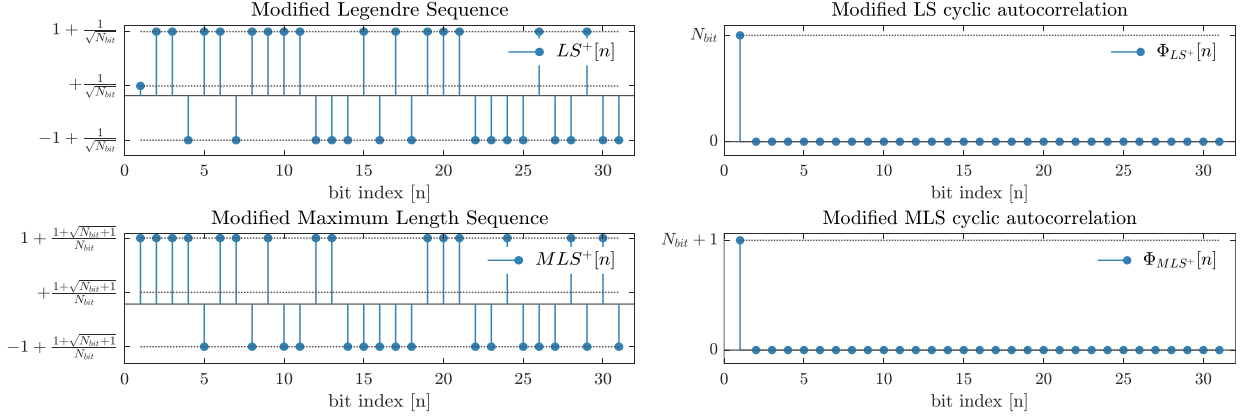

Figure 6: (left) modified $LS^+$ and $MLS^+$ sequences for $N_{bit} = 31$; (right) the corresponding perfect PACFs

As said, there are much more available lengths for $LS$ than $MLS$, so $LS$ are preferable for fine tuning the excitation signals. However, a fully binary excitation is easier to be implemented by just switching ON and OFF a constant heat flux.

So, a further modification is introduced for all the $LS's$ such that $mod(N_{bit}, 4) = 3$. For such sequences, denoted as $LS_{N_{bit}}^{4+}[n]$, it is possible replacing the initial "0" with both "1" or "-1" while maintaining a perfect PACF. This is done as follows:

$$\text{for } N_{bit} \text{ such that } mod(N_{bit}, 4) = 3$$

I. Replace the 1st value $LS_{N_{bit}}[0] = 0 \rightarrow LS_{N_{bit}}[0] = \pm 1$

II. Add a constant bias $LS_{N_{bit}}^{4+}[n] = LS_{N_{bit}}[n] + \frac{\mp 1 + \sqrt{N_{bit}+1}}{N_{bit}}$ (15)

and the PACF is modified as well:

$$\Phi_{LS_{N_{bit}}^{4+}}[n] = \begin{cases} N_{bit} + 1 & \text{for } n = 0 \\ 0 & \text{for } n \neq 0 \end{cases} = (N_{bit} + 1)\delta_{N_{bit}}[n] \quad (16)$$

The demonstration of why Eq.16 is valid only for $mod(N_{bit}, 4) = 3$ lies beyond the scope of this work and it will be discussed in a next paper.

In conclusion, modified $MLS$ and $LS$ sequences assure a perfect PACF, which can be used for PuC applications, as illustrated in Equation 17 and in Figure 7-a.

$$\hat{\delta}(t) = [x * \psi](t) \rightarrow \delta_{N_{bit}}[n] \propto \{\overline{PN} * \overleftarrow{PN}\}_{cyc}[n] \quad (17)$$

The measurement scheme can be thus summarized as follows:
1) the periodic sequence excites the sample;
2) the sample output signal is convolved with the matched filter;
3) the steady-state signal after convolution, i.e. starting from the 2nd period, is the output of the PuC procedure.

*III.2.3 Oversampled PN sequences and their autocorrelation*

Equation 17 above is the discrete-time counterpart of Equation 2, and it shows that cyclic convolution of modified PN sequence assures perfect autocorrelation and hence pulse-compression without any sidelobes can be implemented by using the periodic sequence as excitation signal. However, it is worth highlighting two aspects that impact the practical PuCT implementation:
  a) Equation 17 assures perfect PuC if the sample impulse response lasts less than a sequence PN period or it can be considered negligible after that time.
  b) Equation 17 implicitly considers that in the PuCT procedure illustrated in Figures 1 and 2, the sampling frequency of the output signal, i.e. the thermograms frames-per-second ($FPS$), is synchronous with the PN sequence update rate $f_{Bit}$: $FPS = f_{Bit} = \frac{1}{T_{Bit}}$, where $T_{Bit}$ is the duration of a single sequence bit, see Figure 7-a.

Point a) expresses the time aliasing issue: if a too short sequence period is chosen, the steady state is not reached, and the effect is a time aliasing in the reconstruction of the estimated impulse response. Fortunately, thermal response can be considered negligible after a certain time, so that with a proper choice of the sequence duration, this effect is also negligible.

Point b) instead put a strong constraint, $f_{Bit} = FPS$, in the measurement procedure which is not always optimal. The PN sequence update rate $f_{Bit}$ indeed determines the excited thermal wave bandwidth while $FPS$ can be higher to allow for increasing time-resolution in the reconstructed signals as well as for implementing time-averaging or filtering.

To let the PN-PuCT measurement procedure more flexible, we extend the procedure by introducing an upsampling of the PN sequences, indicated as $PN[n,K]$, where each bit is repeated $K$ times. Practically this corresponds to the condition $FPS = Kf_{Bit}$ so that for each bit of the sequence $K$ thermograms are acquired.

To guarantee that the discrete-time counterpart of Eq. 4 is ensured, the matched filter $\psi PN[n,K]$ of the upsampled sequence is not just the time-reversed replica of the sequence but is obtained by padding with zeros, see Figure 7-b:

$$PN[n,K] = PN[\lfloor n/K \rfloor] \text{ where } \lfloor \cdot \rfloor \text{ is the } floor \text{ function.}$$
$$\psi PN[n,K] = \begin{cases} PN[n/K] & if \ n \equiv 0 \ (\text{mod } K) \\ 0 & elsewhere \end{cases} \quad (18)$$

By exploiting the convolution theorem, it can be seen that:

$$\{PN * \psi PN\}_{cyc}[n] = (N_{bit}+1)\Pi[n,K] = \begin{cases} N_{bit}+1 \ for \ 0 \le n \le K-1 \\ 0 \ for \ K \le n \le KN_{bit} \end{cases} \quad (19)$$

where $\Pi[n,K] = \begin{cases} 1 \ for \ 0 \le n \le K-1 \\ 0 \ elsewhere \end{cases}$ is a discrete-time rectangular pulse with $K$ samples

Equation 19 is thus the discrete-time counterpart of Equation 4, and it assures that after PuC the response to a virtual rectangular pulse of duration $T_{Bit}$ is achieved.

Figure 7 summarizes the two possible PuC strategies resulting using standard and upsampled sequences.

a) $PN - PuCT$ case $FPS = f_{Bit}$

Coded input signal $x[n]$: Periodic sequence $\overline{LS_{11}^{4+}}[n]$

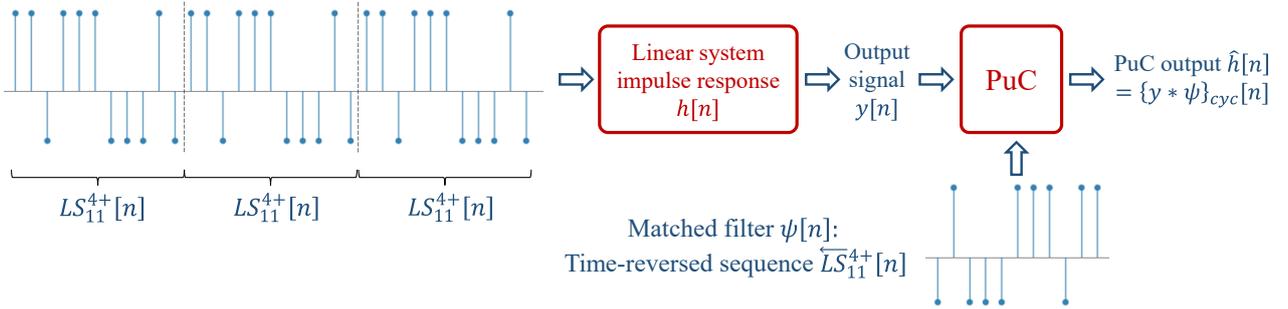

b) $PN - PuCT$ case $FPS = Kf_{Bit}$ ($K=3$)

Coded input signal $x[n]$: Periodic sequence $\overline{LS_{11}^{4+}}[n,3]$

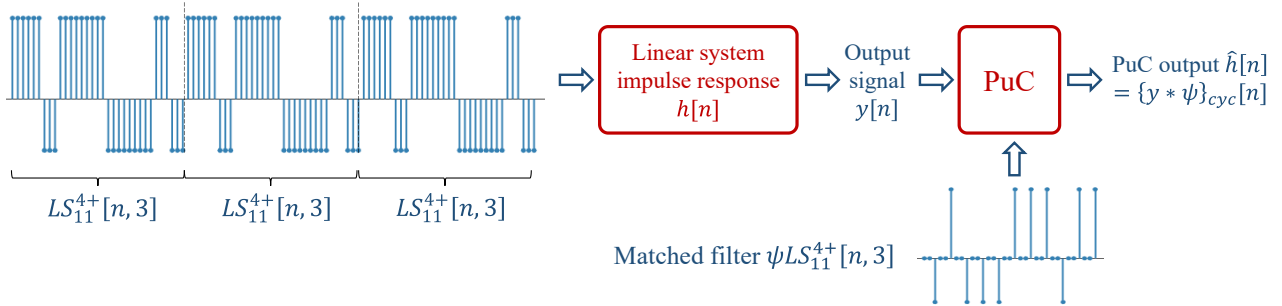

Figure 7: (a) PN-PuCT signals for the case $FPS = f_{Bit}$: the input signal is the periodic repetition of a modified $PN$ sequence with perfect $PACF$ and the matched filter is the time-reversed replica of the sequence (single period); (b) PN-PuCT signals for the case $FPS = Kf_{Bit}$: the input signal is the periodic repetition of a modified $PN$ sequence with perfect $PACF$ which is oversampled $K$ times. The matched filter is the time-reversed replica of the sequence (single period) padded with $K-1$ zeros between each bit.

IV. PSEUDO NOISE PULSE-COMPRESSION THERMOGRAPHY: IMPLEMENTATION ASPECTS

After having introduced all the mathematical aspects of the PN sequences and the related PuC procedures, we can now illustrate how PN-PuCT is practically implemented. This is done by referring to the Figure 8 and explaining step-by-step the various blocks of the diagram. Note that the output signals reported correspond to numerical simulated thermal data for a sound semi-infinite sample where the 1D thermal impulse response defined in [46-47] was used and the thermal diffusivity value was considered was $\alpha = 10^{-6} \left[\frac{m^2}{s}\right]$.

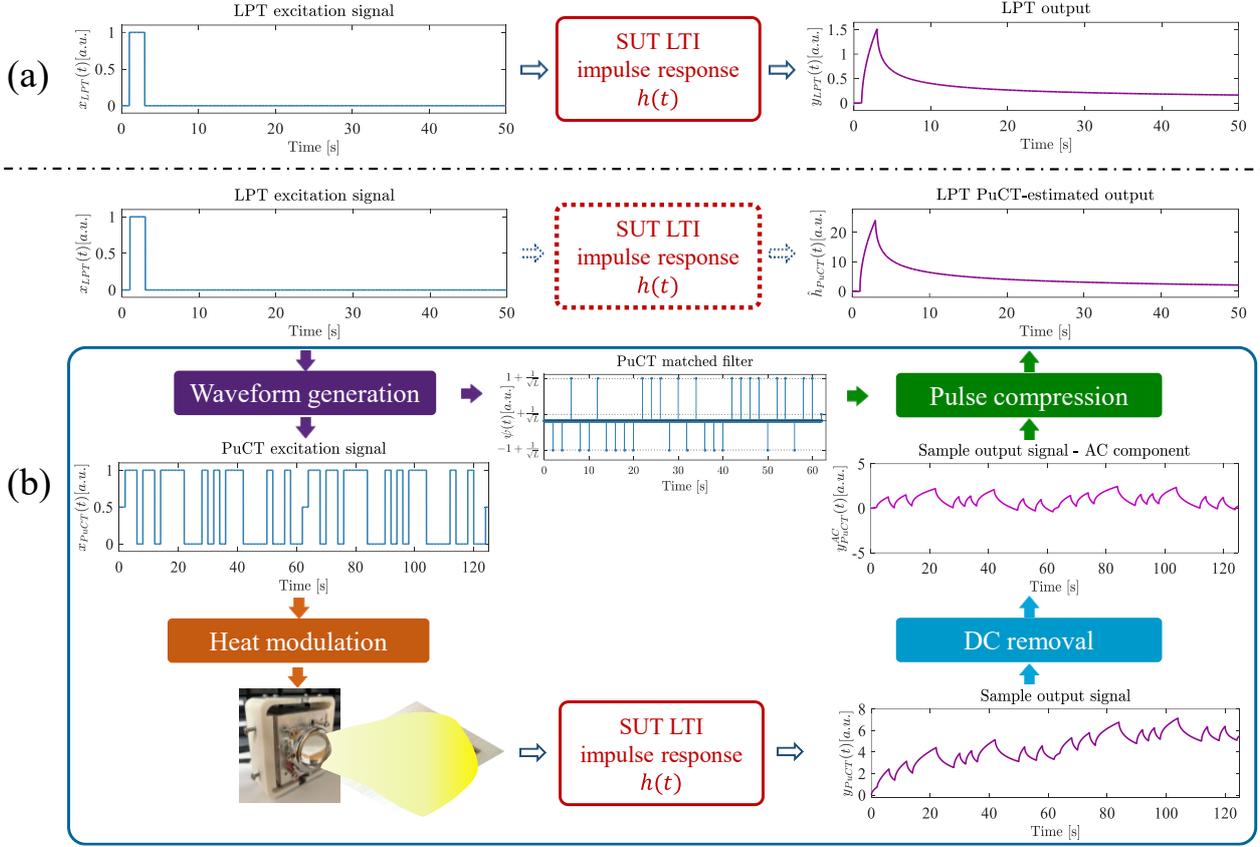

Figure 8: Standard LPT Vs PN-PuCT measurements. Data are simulated by considering the ideal impulse response of a semi-infinite medium of thermal diffusivity $\alpha = 10^{-6} \left[\frac{m^2}{s}\right]$; (a) Typical excitation and output signal for a LPT measure. The pulse duration is 3s and the response is measured for 50 s. (b) PN-PuCT procedure. The same LPT signal of (a) enters the Waveform generation block to define the time-modulation of the heating source. The response of the system to the coded excitation is processed firstly by DC removal and then by the pulse-compression blocks to retrieve the LPT PuCT-estimated output. From an end-user point of view, the whole PuCT procedure can be considered transparent, as depicted by the dashed box representing the linear system. The only effect is the increase in the SNR provided by the PuCT procedure on the LPT PuCT estimated output with respect to the standard LPT measurement.

*IV.1 Heat source modulation waveform*

The heat source modulation waveform is defined starting from the long-pulse excitation $\Pi(t, T_{bit})$ of duration $T_{bit}$ that we want to replace by the PuCT procedure and a PN code of the chosen n° of bits $N_{bit}$. Note that the corresponding modified PN sequence with perfect autocorrelation is used only in the pulse-compression block.

Figure 9 graphically depicts the process for the cases of a 7-bit length maximum length sequence - $MLS_7[n]$- and a 11-bit length Legendre sequence - $LS_{11}[n]$.

Firstly, to obtain the coded waveform $x_{PN}(t, T_{bit})$, the rectangular pulse is repeated $N_{bit}$ times with no delay and its amplitude is modulated by the values of the PN sequences, see Figure 9-a.

$x_{PN}(t, T_{bit})$ lasts for $N_{bit} \times T_{bit} = T_{meas}$ seconds and for $l$ consecutives "1's" or "-1's", the value is kept constant for $l \times T_{bit}$ seconds.

The unipolar excitation modulating the heat source is then obtained by repeating $x_{PN}(t, T_{bit})$ for $N_{per}$ periods, adding a constant value equal to 1, and then multiplying the result for $A/2$, where A is the amplitude of the heat flux, see Figure 9-b. In practical applications $N_{per} = 2$ is enough, however a larger number of repetitions can be used to makes averages.

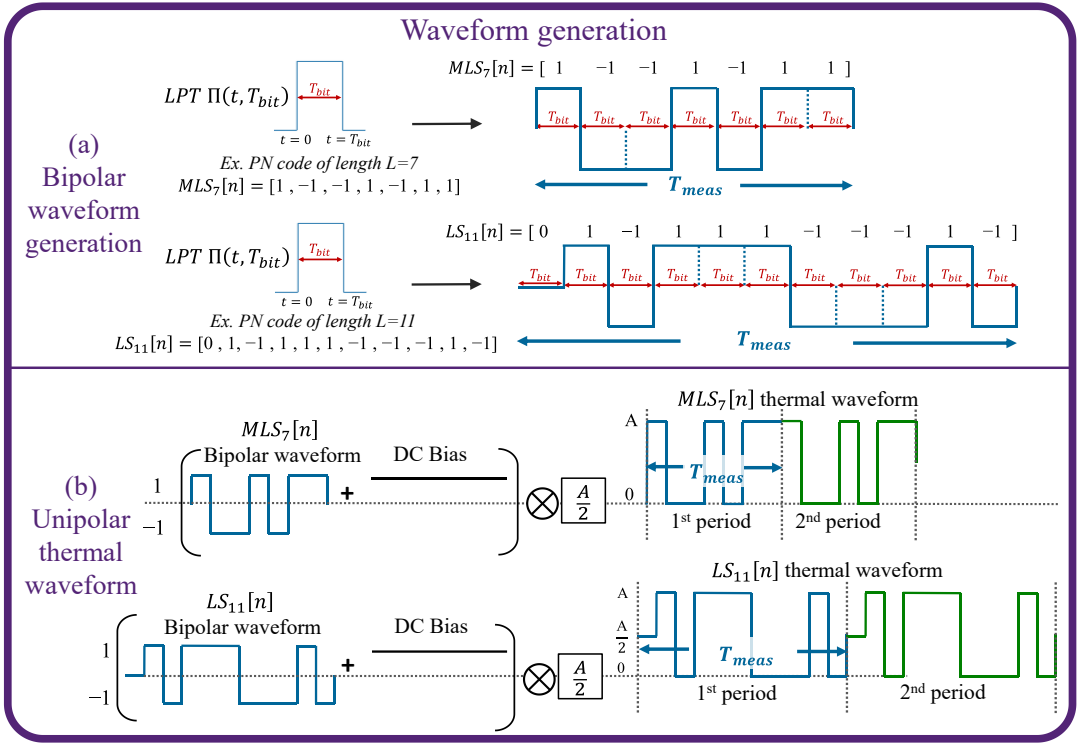

Figure 9. Code generation: a) starting from a long-pulse $\Pi(t, T_{bit})$ of duration $T_{bit}$ and a PN code of the length/n° of bits $N_{bit}$ a bipolar coded waveform of duration $N_{bit} \times T_{bit} = T_{meas}$ is defined; b) the unipolar thermal excitation $x_{TH}(t, T_{bit})$ is obtained by adding a constant value equal to 1 to the bipolar signal, multiplying the result for $A/2$, and then repeating it at least one other time.

More formally, starting from the binary and bipolar numerical sequence $PN[n]$ and the long pulse $\Pi(t, T_{bit})$, the bipolar waveform $x_{PN}(t, T_{bit})$ is defined as:

$$x_{PN}(t, T_{bit}) = \sum_{n=0}^{L-1} PN[n]\Pi(t - nT_{bit}, T_{bit}) \quad \text{for} \quad t \in [0, T_{meas}] \tag{20}$$

while the periodic unipolar thermal excitation containing $N_{per}$ periods of $x_{PN}(t, T_{bit})$ is defined as:

$$\begin{aligned} x_{TH}(t, T_{bit}) &= \sum_{k=0}^{N_{per}-1} \frac{A}{2}\left(x_{PN}(t - kT_{meas}, T_{bit}) + 1\right) = \\ &= \underbrace{\frac{A}{2}\left(\theta(t) - \theta(t - N_{per}T_{meas})\right)}_{\text{DC component } x_{TH}^{DC}(t,T)} + \underbrace{\sum_{k=0}^{N_{per}-1} \frac{A}{2} x_{PN}(t - kT_{meas}, T_{bit})}_{\text{AC component } x_{TH}^{AC}(t,T)} \end{aligned} \tag{21}$$

where the last expression highlights the interpretation of the excitation as a sum of a "DC" term $x_{TH}^{DC}(t, T_{bit})$ and a "AC" one $x_{TH}^{AC}(t, T_{bit})$, both lasting for $N_{per} \times T_{meas}$ seconds, such that $x_{TH}(t, T_{bit}) = x_{TH}^{DC}(t, T_{bit}) + x_{TH}^{AC}(t, T_{bit})$.

In practice, if a pure binary sequence is used, e.g. any MLS or a modified LS, the thermal waveform can be implemented by just switching on and off a constant heat thermal source, thus simplifying the setup with respect to the use of a frequency modulated signal which requires an analog modulation of the heat power.

The product $N_{bit} \times T_{bit} = T_{meas}$ is the duration of a single period of the modulation waveform and it is also the duration of the LPT output obtained at the end of the PuCT procedure.

From a NDE perspective, $T_{meas}$ must be long enough to gather all the information needed for the sample characterization. From a mathematical point of view, $T_{meas}$ should be longer than the expected LPT output, but in practice such constraint can be relaxed choosing $T_{meas}$ long enough so that the expected LPT output at $t \geq T_{meas}$ is close to original state.

$T_{meas}$ and $T_{bit}$ must be therefore defined by starting from the characteristic of the inspected sample and considering the a priori information about expected types and location of the defects, if any. Once defined $T_{bit}$ and the minimum $T_{meas}$ value ($T_{meas}^0$), the code length $N_{bit}$ can assume any value satisfying $N_{bit} \times T_{bit} \geq T_{meas}^0$. Under this condition, $N_{bit}$ can be increased arbitrarily to increase the SNR proportionally.

The output of the sample to the thermal excitation is a sequence of thermograms acquired at the rate $FPS$. Note that $T_{bit}$ and the $FPS$ cannot be independent, as explained in Subsection III.2.3. To correctly implement the PuCT procedure, $T_{bit}$ must be equal or a multiple of the time interval $\delta t = \frac{1}{FPS}$ between two consecutive frames, that is, for each bit of the sequence an integer number of thermograms must be collected: $T_{bit} = \frac{K}{FPS}$ with $K \in \mathbb{N}^+$, where $K$ is the oversampling factor. Note also that with the sampling frequency equal to $FPS$, $\Pi(t, T_{bit})$ is represented by a sequence of $K$ samples as expressed in Equation 19.

The sequence of thermograms $\mathbf{y}[n\delta t]$ can be seen as a 2D set of discrete-time signals, corresponding to the time-trend of the pixels' intensity, where $y_{j_x,j_y}[n\delta t]$ is the time-trend of the $(j_x, j_y)$ pixel. Due to the linearity assumption, $y_{j_x,j_y}[n\delta t]$ can be decomposed in a DC and AC component too, as follows:

$$y_{j_x,j_y}[n\delta t] = \underbrace{y_{j_x,j_y}^{DC}[n\delta t]}_{\text{due to } x_{TH}^{DC}(t,T)} + \underbrace{y_{j_x,j_y}^{AC}[n\delta t]}_{\text{due to } x_{TH}^{AC}(t,T)} \qquad (23)$$

*IV.2 DC removal*

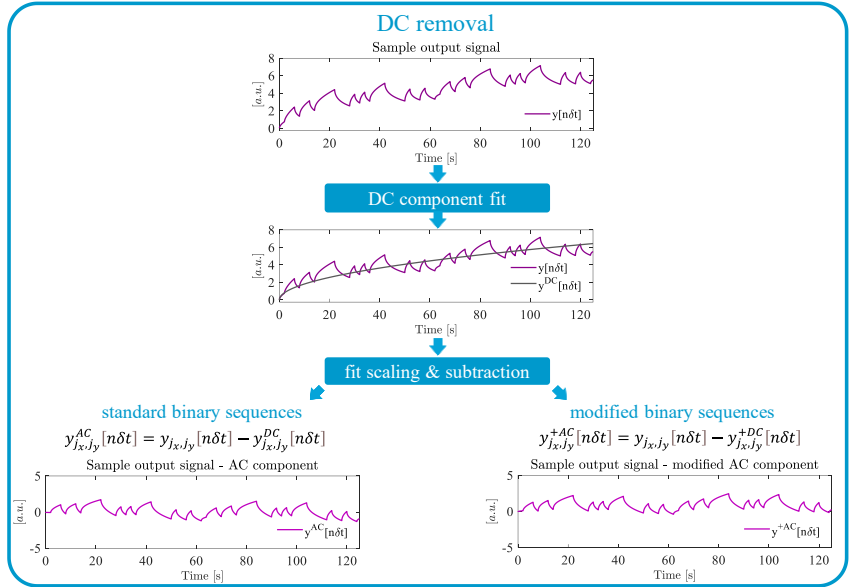

Figure 10. DC removal: for each pixel $(j_x, j_y)$ the time-trend of the intensity $y_{j_x,j_y}[n\delta t]$ is fitted by a polynomial function $y_{DC}(t) = a_1 t + a_2 t^{0.75} + a_3 t^{0.5}$ and then the fitted curve $y_{j_x,j_y}^{DC}[n\delta t]$ is: (left, standard sequences case) subtracted to $y_{j_x,j_y}[n\delta t]$ to obtain the AC component $y_{j_x,j_y}^{AC}[n\delta t]$, (right, modified sequences case) multiplied by a factor $(1 - bias)$ and then subtracted to $y_{j_x,j_y}[n\delta t]$ to obtain the output signal $y_{j_x,j_y}^{+AC}[n\delta t]$.

The DC removal step consists in estimating the contribute to the sample output signal due to the DC component of the thermal excitation and remove it before implementing the PuC step or other processing. This procedure is now quite a standard, see for instance [19, 28]. However, in the present case also the DC removal operation must be slightly modified. This is since the pulse-compression procedures that we want to implement are based on modified sequences, see Figure 7, and all the modified sequences have a not-vanishing DC bias. Thus, we do not subtract the whole DC component, but a scaled version that considers the DC bias in the modified MLS and LS sequences.

Figure 10.a graphically depicts how the DC removal works in both the cases of standard, i.e. of "+1" and "-1" sequences, and modified ones. For each pixel $(j_x, j_y)$ the time-trend of the intensity $y_{j_x,j_y}[n\delta t]$ is fitted by a polynomial function $y_{DC}(t) = a_1 t + a_2 t^{0.75} + a_3 t^{0.5}$, where the fit coefficients $a_k$ are constrained to be positive or null, and then the fitted curve $y_{j_x,j_y}^{DC}[n\delta t]$ is:

a) subtracted to $y_{j_x,j_y}[n\delta t]$ to obtain the AC component $y_{j_x,j_y}^{AC}[n\delta t]$, see Eq.23.
b) multiplied by a factor $(1 - bias)$ and then subtracted to $y_{j_x,j_y}[n\delta t]$ to obtain the output signal $y_{j_x,j_y}^{+AC}[n\delta t]$ corresponding to a modified sequence as excitation.

To better understand this, we can rewrite Eq. 23 as follows:

$$y_{j_x,j_y}[n\delta t] = y_{j_x,j_y}^{DC}[n\delta t] + y_{j_x,j_y}^{aC}[n\delta t] = \qquad (24)$$

$$= \left\{y^{DC}_{j_x,j_y}[n\delta t] * (1 - bias)\right\} + \underbrace{\left\{y^{AC}_{j_x,j_y}[n\delta t] + y^{DC}_{j_x,j_y}[n\delta t] * bias\right\}}_{\text{due to modified sequences excitation}} = y^{+DC}_{j_x,j_y}[n\delta t] + y^{+AC}_{j_x,j_y}[n\delta t]$$

where $bias = \frac{1+\sqrt{N_{bit}+1}}{N_{bit}}$ for $MLS^+_{N_{bit}}$; $bias = \frac{1}{\sqrt{N_{bit}}}$ for $LS^+_{N_{bit}}$; $bias = \frac{\mp 1+\sqrt{N_{bit}+1}}{N_{bit}}$ for $LS^{4+}_{N_{bit}}$

*IV.3 Pulse compression*

After the DC removal, the PuC step is implemented. Being based on cyclic convolution, for each pixel the signal $y^{+AC}_{j_x,j_y}[n\delta t]$ corresponding to 2 or more excitation periods is convolved with the matched filter and from the steady-state of the convolution output the true PuCT result $\hat{h}_{j_x,j_y}[n\delta t]$ is extracted. As explained in Subsection III.2.3 and in Figure 7, the matched filter is equal to $\overleftarrow{PN}[n\delta t]$ for $FPS = f_{Bit} = \frac{1}{T_{Bit}}$, and equal to $\psi PN[n\delta t, K]$ $FPS = Kf_{Bit} = \frac{K}{T_{Bit}}$. Figure 11 illustrates the pulse compression step for the case of a modified Legendre sequence of length 31, $LS^{4+}_{31}[n]$, with 3 periods of excitation.

It is worth highlighting that we are assuming $T_{meas}$ long enough so that the sample impulse response for each pixel is null or almost null for $t = T_{meas}$. However, in some experimental cases, e.g. for very thick samples or for very small thermal conductivity values, this can a strong assumption. This can lead to a non-perfect periodic response of the cyclic convolution and thus to a small deviation of the reconstructed response from the ideal one. Similar arguments apply also for the DC removal, as the polynomial fitting is just an approximation of the ideal unknown behavior. However, these drawbacks of the technique are more than counterbalanced by the great advantage in terms of SNR that the PN-PuCT provide in presence of small heating power and/or short duration of the pulse excitation.

The experimental results reported in the next Section will demonstrate the effectiveness of the procedure for a range of $T_{Bit}$ duration, sequence lengths, etc..

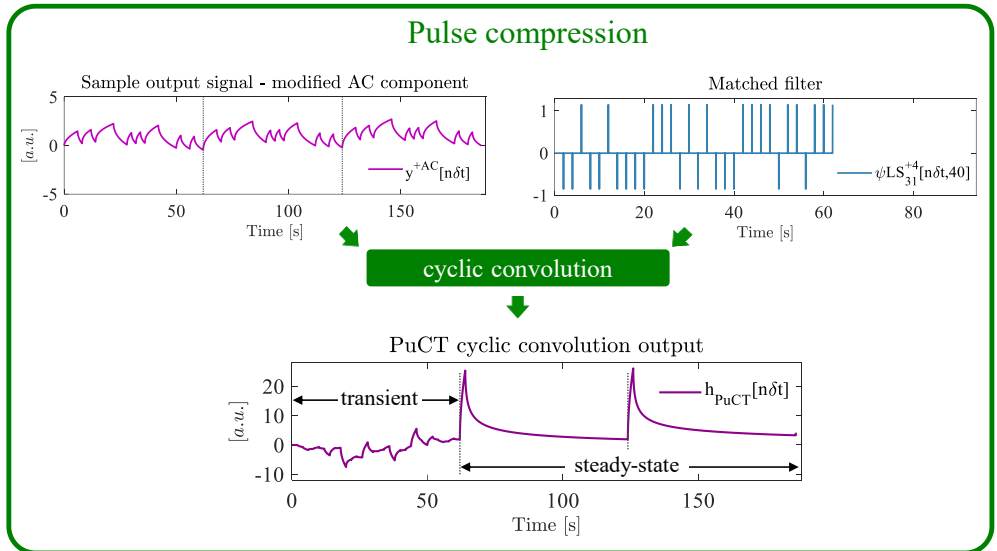

Figure 11. Pulse compression: for each pixel $(j_x, j_y)$ the output $y^{+AC}_{j_x,j_y}[n\delta t]$ of the DC removal block is convolved with the matched filter $\psi[n\delta t]$. The convolution output is characterized by a first transient period followed by a periodic steady state. Each period of the steady state is the PuCT output $\hat{h}_{j_x,j_y}[n\delta t]$.

## V. EXPERIMENTAL RESULTS

The PN-PuCT procedure described above has been tested and applied by using different PN codes, for both conditions $FPS = f_{bit}$ and $FPS = Kf_{bit}$, with a variety of samples, and by using different heating sources (LEDs, halogen lamps, induction heating, etc..). Without being exhaustive, in the present paper we illustrate some of these results that give an insight on the main features and benefits of the procedure.

In all the cases reported, standard and modified Legendre sequences were used. The heating source consisted of a set of high-power LED chips, with total maximum power of $320W$ driven by a programmable voltage generator (TDK Lambda GEN 750W) or by a fixed voltage generator followed by a MOSFET acting as a switch. The MOSFET-based solution was used to increase switching rapidity, allowing very small values of $T_{bit}$ to be selected (up to $\sim 1ms$), portability, and for reducing the costs of the heating system. More details on the LED-based heating system will be the subject of a further publication. The heat source modulation waveform was generated by means of an arbitrary waveform generator (AWG), which also provide the trigger/external clock to the IR camera for assuring the synchronization between the start of the heat excitation and the acquisition of the thermograms at a given constant FPS value. A National Instruments myDAQ board or a Tie-Pie HS5 were used as AWG while the IR camera used was a Xenics Onca-MWIR-InSb with 320 × 256 pixels. Different samples-under-test (SUTs) were tested, having different thermal properties, defect types and structures.

Figure 12 contains a sketch of the typical experimental setup (a), and a sketch of the benchmark sample SUT1 used to verify the soundness of the procedure (b). The benchmark SUT has been selected since for it pulsed-thermography measurements as well data obtained with the same setup of the present paper by using Barker codes and frequency-modulated excitations were reported in [19] and [20], so the reader is referred to those works for a comparison of the present results with previous ones.

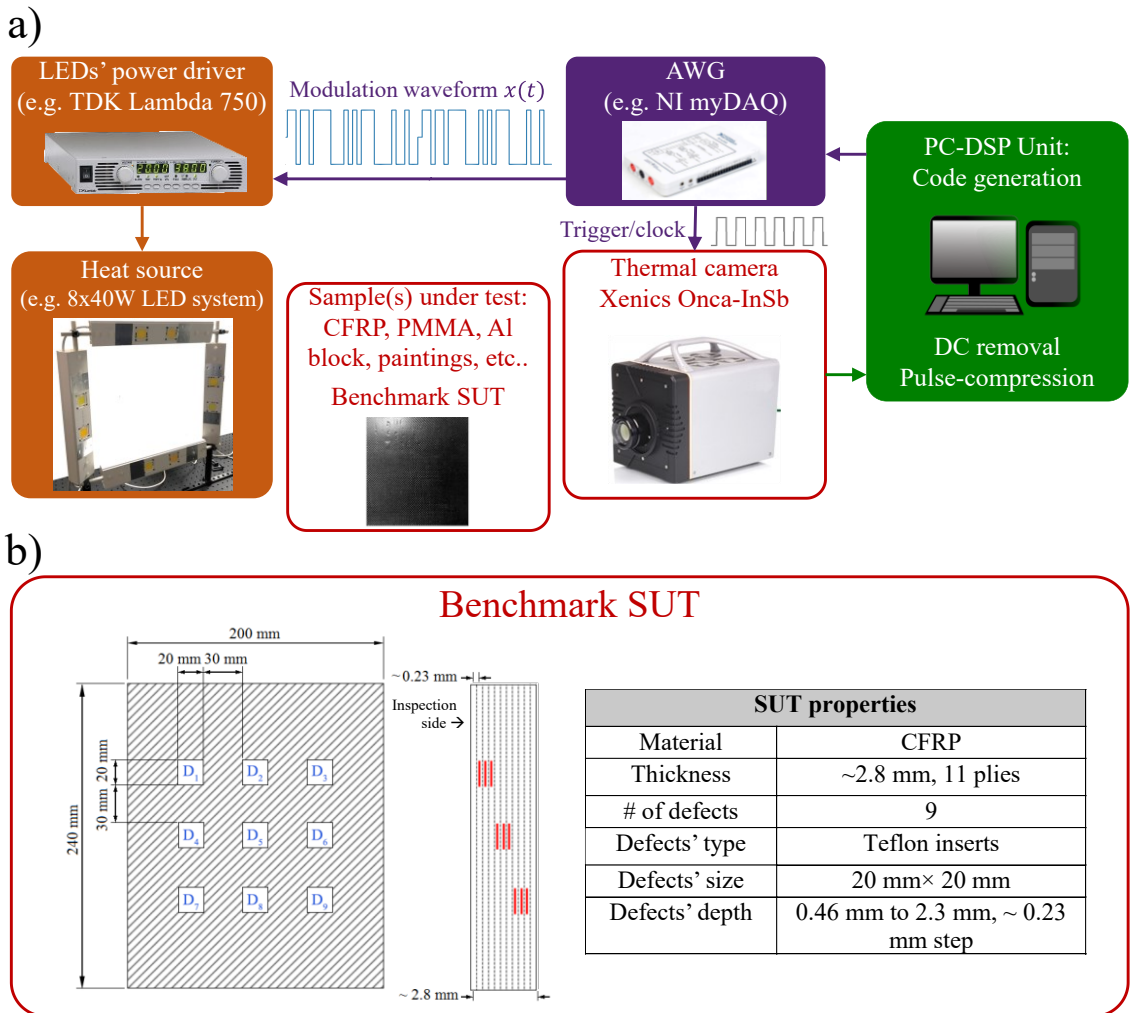

Figure 12. a) typical experimental setup block diagram: a PC-DSP unit manage the hardware and inputs the arbitrary waveform generator (AWG) which generates the modulation waveform, see Section 4.1, and the trigger/clock signal of the IR camera. The modulation waveform is the input signal of the LEDS' power driving system. The LEDs system illuminates, and thus heats, the SUT and the thermal camera collects a sequence of thermograms during all the heating time. This sequence is then processed (DC removal + pulse compression), to return the PuCT output. (b) sketch and properties of the benchmark sample used to verify the soundness of the procedure.

*V.1 Comparison of different $T_{bit}$ values*

The main characteristic of the present PN-PuCT procedure is that at the end of the procedure we obtain the response to a virtual LPT excitation of duration equal to $T_{bit}$ and energy equal to half of the modulation waveform. We started therefore verifying this aspect.2

We did four experiments by using different values of $T_{bit}$, namely $T_{bit} = 0.5s, 1s, 1.4s, 1.9s$ and to have a fair comparison in terms of SNR we used different lengths of $LS$s, respectively $N_{bit} = 61, 31, 23, 17$ so as to have almost constant the product $T_{bit} \times N_{bit}$, which is proportional to the excitation energy. In all the experiments the thermograms acquisition rate was $FPS = 40$, so the oversampling factor was respectively $K = 20, 40, 56, 76$. For each measurement, we used $N_{per} = 2$ periods of the PN excitation and collected $N_{per} \times K \times N_{bit}$ frames. We selected for comparison three pixels over the sample, one corresponding to a sound point, denoted by $S_1$, one in correspondence of the defect #5, denoted by $D_5$, and one over the shallowest defect, denoted by $D_1$. These pixels are marked on the image on the left of Figure 13 that visualizes the PuCT output $\hat{h}_{PuCT}$ at $t = 6s$.

Firstly, we verified the various steps of the procedure for the case $N_{Bit} = 61$. The subplots on the right of Figure 13 show the various signals introduced in Subsections V.2 and V.3 for the sound point $S_1$. On top is depicted the acquired signal $y[n\delta t]$ together with the fit of the DC component $y^{DC}[n\delta t]$ and the scaled DC component $y^{+DC}[n\delta t]$. In the middle the signal $y^{+AC}[n\delta t]$ after the DC removal is shown, and on the bottom the PuCT output signal $\hat{h}_{PuCT}[n\delta t]$ is reported. It can be seen that $\hat{h}_{PuCT}[n\delta t]$ behaves like a response to a pulse of duration $T_{bit} = 0.5s$: the temperature increases from $t = 0s$ up to $t = 0.5s$ and then the cooling starts. The cooling is monotonic, and there are not any visible sidelobes in the reconstructed LPT output.

Then we repeated the PuCT procedure for the four measurements. Figure 14 reports the DC removal step for all the $L$ and $T_{bit}$ values for the sound point $S_1$: it can be appreciated how the same fitting function works quite well in all the cases and how the dynamics of the $y^{+AC}[n\delta t]$ signals changes depending on the $T_{bit}$ duration. This also gives an insight on how to chose $T_{bit}$ to maximize the trade-off between excited bandwidth and SNR: actually the SNR after PuCT is determined by the energy of $y^{+AC}[n\delta t]$, so from Figure 14 it can be argued that the maximum achievable SNR is expected for $T_{bit} = 1.9s$, even if the energy of $y^{+DC}[n\delta t]$ is almost the same for all the $T_{bit}$'s. In general, if after the DC removal the signal is very weak, the SNR will be poor as well. If a large bandwidth, i.e. a small $T_{bit}$, is needed to increase the time resolution, then a longer code can be used to increase the SNR.

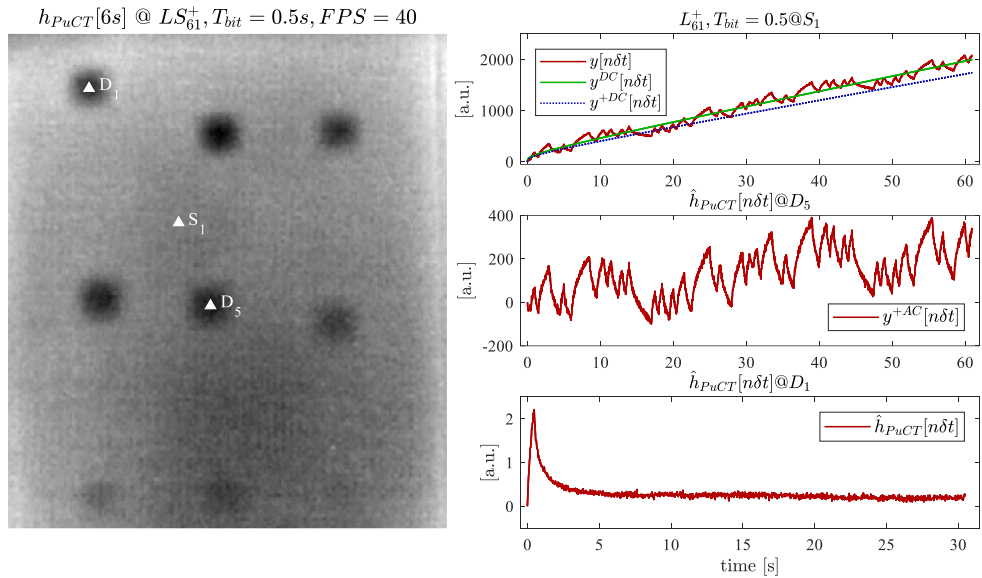

Figure 13. Signals involved in the various steps of the PN-PuCT procedure: (top-right) the sample response time-trend $y[n\delta t]$ to the thermal excitation (red solid line) plotted together with the fit of the DC component $y^{DC}[n\delta t]$ (green solid line) and the scaled DC component $y^{+DC}[n\delta t]$ that is subtracted before applying the pulse compression; (middle) the resulting $y^{+AC}[n\delta t]$ signal after the DC removal step; (bottom) the output $\hat{h}_{PuCT}[n\delta t]$ of the PuC block.

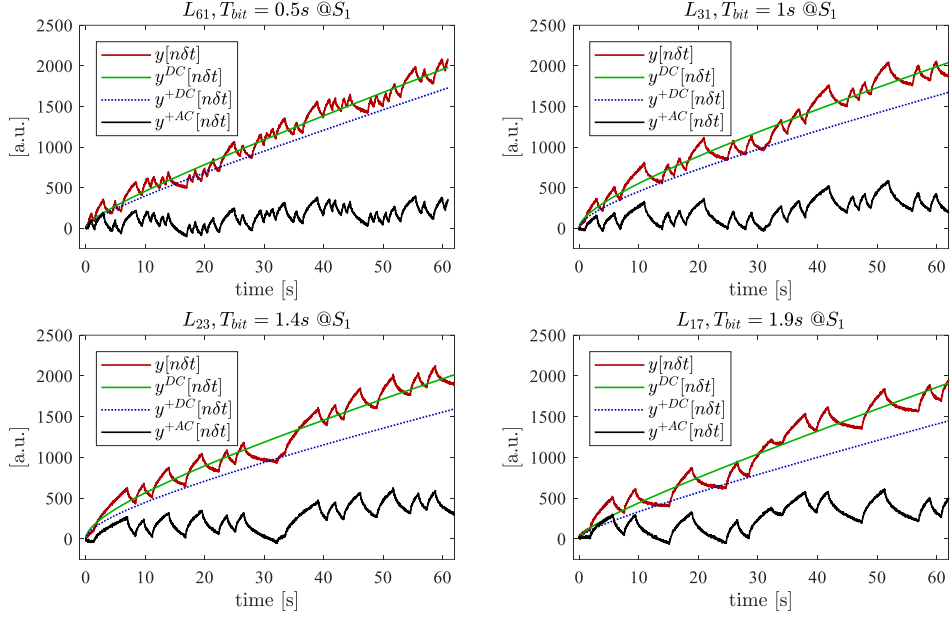

Figure 14. DC removal step for the four measurements: the signals $y[n\delta t]$, $y^{DC}[n\delta t]$, $y^{+DC}[n\delta t]$ and $y^{+AC}[n\delta t]$ are shown for the sound point $S_1$.

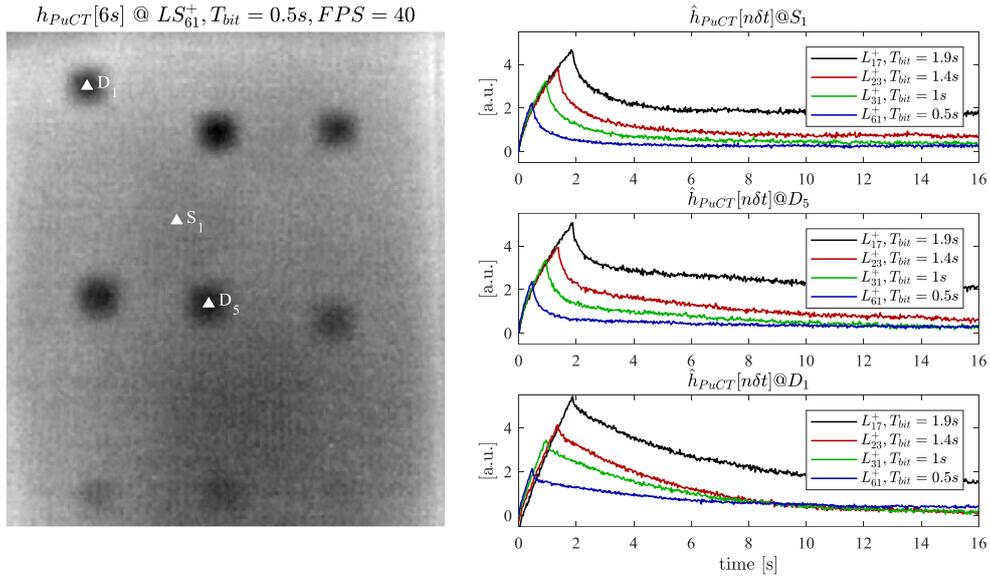

Figure 15. Comparison of the $\hat{h}_{PuCT}[n\delta t]$ obtained for different values of $T_{bit}$ and $N_{bit}$. The plots on the right show the $\hat{h}_{PuCT}$'s corresponding to three different points, $S_1$, $D_5$ and $D_1$, marked in the image on the left. The plots were normalized by the length of the code $L$. It can be seen that: i) after the normalization, the various curves are almost perfectly superimposed in the initial heating interval; ii) there are not oscillations of the temperature during the cooling neither any other sign of sidelobes effects after the PuC; iii) both the sound and the defects' curves show the typical behaviour of the LPT response.

Figure 15 summarizes and compares the results of the PuCT for the four measurements, after having normalized the $\hat{h}_{PuCT}[n\delta t]$ for the code length $L$ values. It can be seen that: i) the various curves after the pulse compression are almost perfectly superimposed in the initial heating stage as expected for measurements done with LPT signals of varying duration collected with the same initial conditions; ii) there are not oscillations of the temperature during the cooling neither any other sign of sidelobes effects after the PuC; iii) both the sound and the defects' curves show the typical behavior of the LPT response.

Looking at the plots in the right side of Figure 15 is indeed not possible to distinguish if they are obtained from LPT experiments or after PN-PuCT ones, and this is the main goal of the present paper.

To further corroborate this results, in Figure 16 the time signals $\hat{h}_{PuCT}[n\delta t]@D_k$ over the nine defects are depicted together with $\hat{h}_{PuCT}^{REF}[n\delta t]$, which is the mean of $\hat{h}_{PuCT}$ over the whole sample.

The time trends of figure 16 look very like those of LPT experiments, but the SNR levels are much higher than those achievable with long pulse excitations with the same peak power.

To have a quick overview of the defects' detection capability, in Figure 17 some images of $\hat{h}_{PuCT}[n\delta t]$ at different $n\delta t$ values are reported for the four measurements.

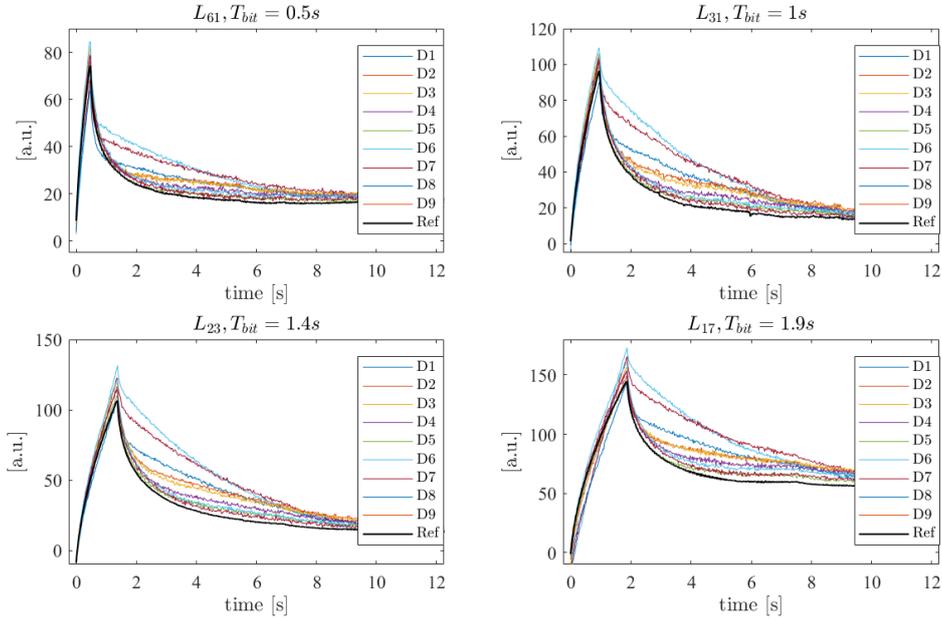

Figure 16. Plots of the time signals $\hat{h}_{PuCT}[n\delta t]@D_k$ over the nine defects (D1-D9) for the different values of $T_{bit}$ and $N_{bit}$. The black line in each plot represents $\hat{h}_{PuCT}^{REF}[n\delta t]$, which is the mean of $\hat{h}_{PuCT}$ over the whole sample.

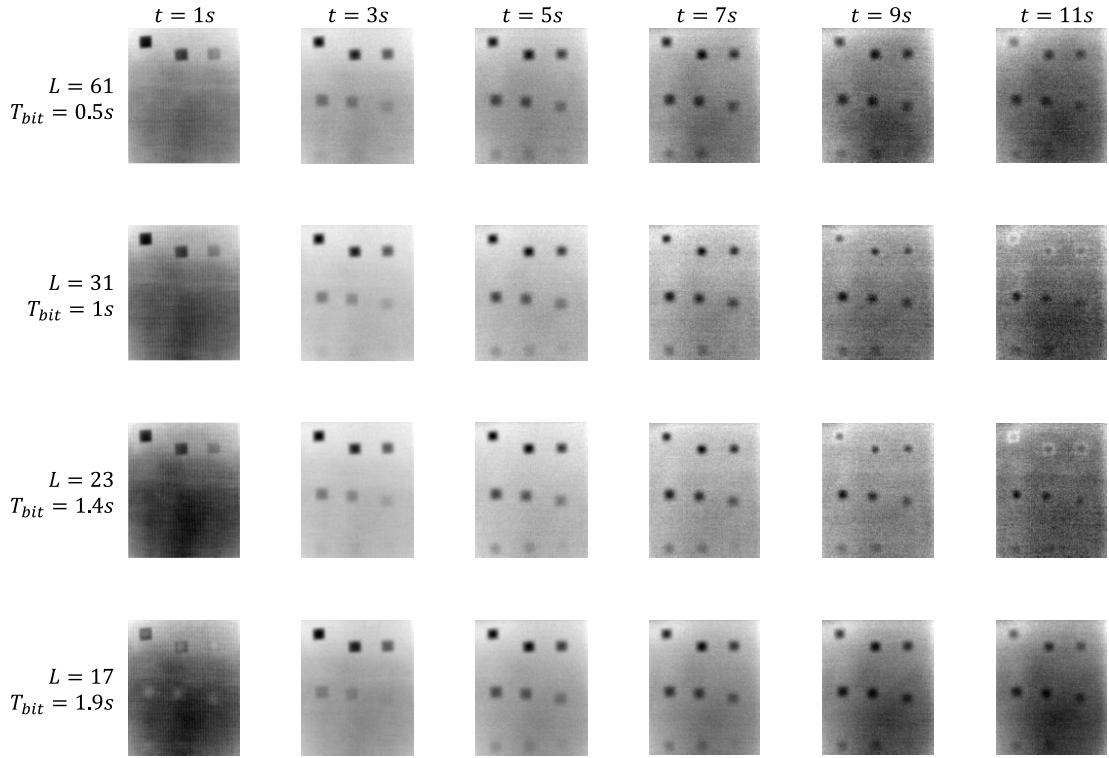

Figure 17. Images of $\hat{h}_{PuCT}[n\delta t]$ at different $n\delta t$ values are reported for the four measurements. The defects D1-D8 can be clearly seen for all the $T_{Bit}$ values tested, even if the SNR for deeper defects seems to be larger for larger $T_{Bit}$, as expected.

The defects D1-D8 can be clearly seen for all the $T_{Bit}$ values tested, even if the SNR for deeper defects seems to be larger for larger $T_{Bit}$, as expected. These results are perfectly in line with those reported in [19, Figures 10 and 14]. D9 is also visible, even if with a low SNR, improving the results reported in 19. Note that the defects' detection capability and the SNR can be increased by post-processing of PuCT images, for example by visualizing the time-phase features, as done in [3,48], or by applying a moving average filter on the $\hat{h}_{PuCT}$. However, we do not want here to focus on how to maximize the SNR and/or the defects' detection, but on the realization of a pulse compression thermography procedure "sidelobes free".

### V.2 Case $FPS = f_{Bit}$

Another property of the PN-PuCT procedure that we tested is the possibility to work with $FPS = f_{Bit}$, that is, collecting only one thermogram for each bit of the excitation, whichever is the $T_{Bit}$ value. So, we repeated the PuC procedure for the four measurements described above on subsampled sequences to process only $N_{per} \times L$ frames where $N_{per} = 2$ is the number of periods.

Like Figure 14, Figure 18 reports the time trend of $y[n\delta t]$, $y^{DC}[n\delta t]$, $y^{+DC}[n\delta t]$ and $y^{+AC}[n\delta t]$ for the 4 measurements for the sound point $S_1$. But differently from the curves reported in Figure 14, here the signals have a different $FPS$ and hence $\delta t$, which is equal to $T_{Bit}$, as highlighted by the dotted curves.

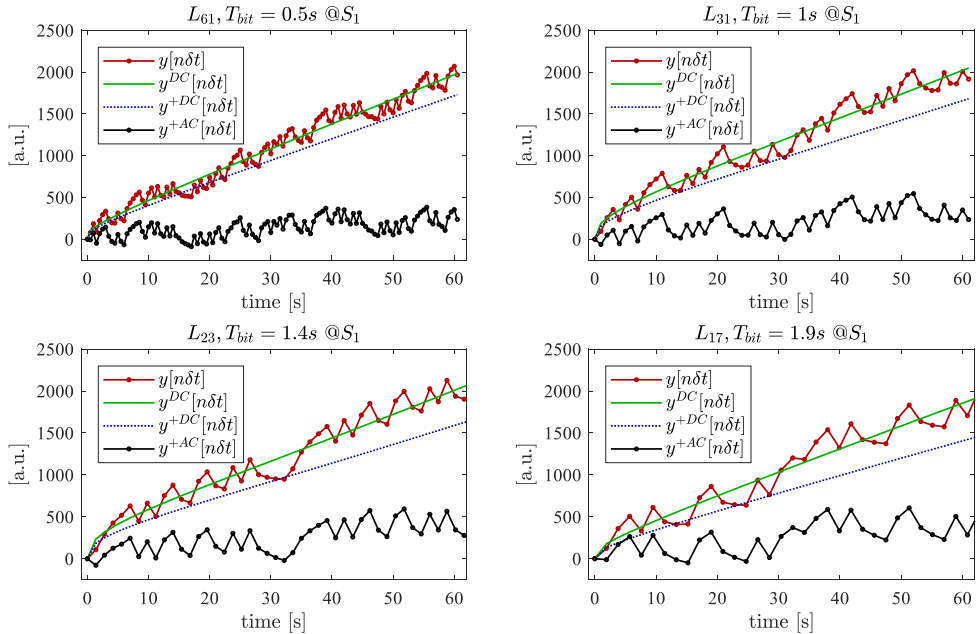

Figure 18. Case $FPS = f_{Bit}$: DC removal step for the four measurements: the signals $y[n\delta t]$, $y^{DC}[n\delta t]$, $y^{+DC}[n\delta t]$ and $y^{+AC}[n\delta t]$ are shown for the sound point $S_1$.

As for Figure 16, Figure 19 reports the $\hat{h}_{PuCT}[n\delta t]'s$ reconstructed over the nine defects (D1-D9) for the different values of $T_{bit}$ and $N_{bit}$. As before, the black line in each plot represents the mean of $\hat{h}_{PuCT}$ over the whole sample, $\hat{h}_{PuCT}^{REF}[n\delta t]$.

It's interesting to notice that the trends look very similar despite the different sampling rate, and that, even by processing a few thermograms, especially in the cases $L = 23$ and $L = 17$, it is possible to differentiate defects from sound points.

This is further verified in Figure 20 by imaging the $\hat{h}_{PuCT}[n\delta t]'s$ at different $t$ values for the four measurements.

Despite the smaller number of thermograms processed, most of the defects are still clearly visible for all the $T_{Bit}$ values tested, and D8 and D9 are barely visible.

These results are quite good, especially for $L = 17$ the number of thermograms processed is very small, equal to 34, 76 time smaller than in the case of figure 17, nonetheless D8 is clearly visible and D9 barely.

This makes this procedure an effective compromise between SNR optimization and computational time and burden: in some applications, like inline test, it could be possible to preprocess the acquired data at $FPS = $

$f_{Bit}$ to have a first glance of the quality of the sample and use this preview to decide about doing or not the analysis at the higher $FPS$ values.

But the case $FPS = f_{Bit}$ is not good only for such goal,

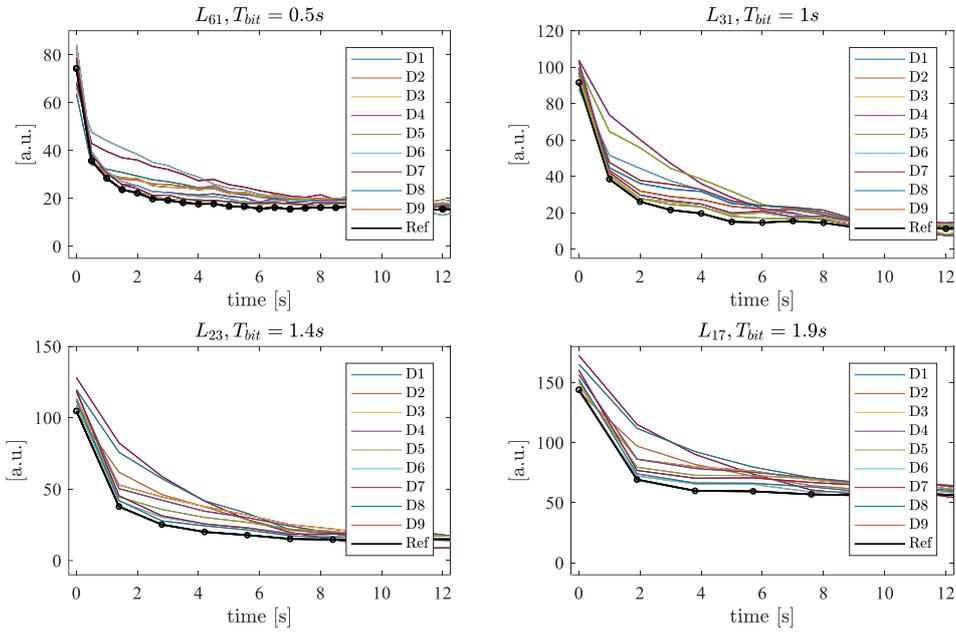

Figure 19 Case $FPS = f_{Bit}$: Plots of $\hat{h}_{PuCT}[n\delta t]@D_k$ over the nine defects (D1-D9) for the different values of $T_{bit}$ and $N_{bit}$. The black line in each plot represents $\hat{h}_{PuCT}^{REF}[n\delta t]$, which is the mean of $\hat{h}_{PuCT}$ over the whole sample

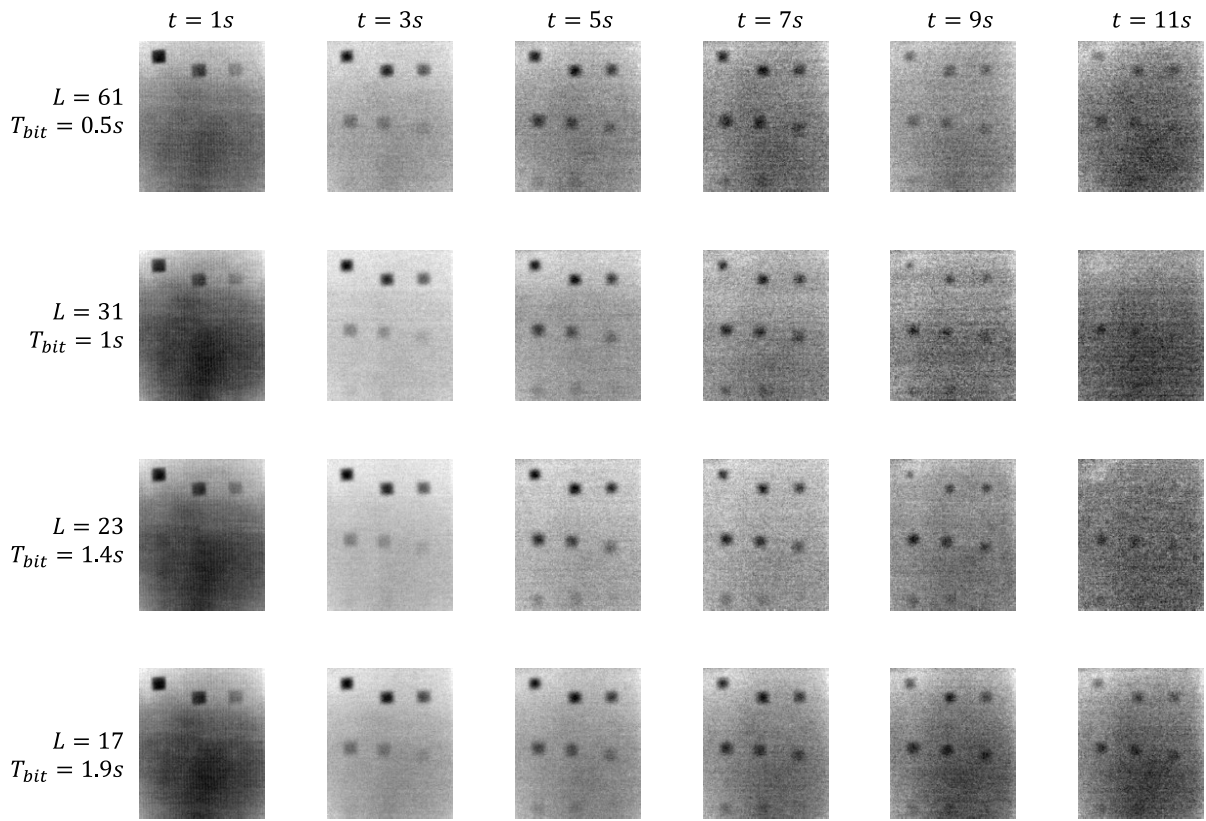

Figure 20 Case $FPS = f_{Bit}$: Images of $\hat{h}_{PuCT}[n\delta t]$ at different $n\delta t$ values are reported for the four measurements. The defects D1-D8 can be clearly seen for all the $T_{Bit}$ values tested, even if the SNR for deeper defects seems to be larger for larger $T_{Bit}$, as expected.

## VI. Conclsions

The theory of a novel pulse-compression thermography procedure has been described and experimentally demonstrated. The procedure can completely suppress the sidelobes affecting all the previous pulse-compression thermography / thermal wave radar results reported in the literature.

Furthermore, the proposed procedure outputs the response of a virtual rectangular pulse of duration defined by the code parameters, thus retrieving the results of a Long-pulse thermography but with increased SNR.

This makes also possible to tune the duration of the virtual pulse to cope with different materials and applications.

Furthermore, the implementation of the procedure with $FPS = f_{bit}$ provide an effective way to optimize the trade-off between SNR enhancement and computational time/resources.

The present procedure is suitable for the use with laser-thermography, where the modulation time, i.e. $T_{bit}$, could be very short, and with any active thermography schemes in which the heating can be easily switched on and off in a controlled way, e.g. induction heating, vibro-acoustic thermography, microwave thermography, air-jet thermography, etc. and finding applications in nondestructive evaluation, medical diagnostics, heritage science, etc.

Results of the application of the procedure to such scheme will be subject of future investigations.

## Acknowledgement

This work was partially supported by: the POS RADIOAMICA project funded by the Italian Minister of Health (CUP: H53C22000650006), the project "START - SusTainable dAta-dRiven manufacTuring", CUP: B29J23000170005, F/310087/01- 05/X56; funded by MIMIT, DM 31/12/2021; and the Next Generation EU - Italian NRRP, Mission 4, Component 2, Investment 1.5, call for the creation and strengthening of 'Innovation Ecosystems', building 'Territorial R&D Leaders' (Directorial Decree n. 2021/3277) - project Tech4You - Technologies for climate change adaptation and quality of life improvement, n. ECS0000009. This work reflects only the authors' views and opinions, neither the Ministry for University and Research nor the European Commission can be considered responsible for them.

## References

[1] Tuli, S., & Mulaveesala, R. (2005). Defect detection by pulse compression in frequency modulated thermal wave imaging. *Quantitative InfraRed Thermography Journal*, *2*(1), 41-54.

[2] Tabatabaei, N., Mandelis, A., & Amaechi, B. T. (2011). Thermophotonic radar imaging: An emissivity-normalized modality with advantages over phase lock-in thermography. *Applied Physics Letters*, *98*(16).

[3] Tabatabaei, N., & Mandelis, A. (2009). Thermal-wave radar: A novel subsurface imaging modality with extended depth-resolution dynamic range. *Review of Scientific Instruments*, *80*(3), 034902.

[4] Almond, D. P., & Lau, S. K. (1994). Defect sizing by transient thermography. I. An analytical treatment. *Journal of Physics D: Applied Physics*, *27*(5), 1063.

[5] Maldague, X., & Marinetti, S. (1996). Pulse phase infrared thermography. *Journal of applied physics*, *79*(5), 2694-2698.

[6] Shepard, S. M. (2001, March). Advances in pulsed thermography. In *Thermosense XXIII* (Vol. 4360, pp. 511-515). SPIE.

[7] Balageas, D. L. (2012). Defense and illustration of time-resolved pulsed thermography for NDE. *Quantitative InfraRed Thermography Journal*, *9*(1), 3-32.

[8] Wu, D., & Busse, G. (1998). Lock-in thermography for nondestructive evaluation of materials. Revue générale de thermique, 37(8), 693-703.

[9] Pickering, S., & Almond, D. (2008). Matched excitation energy comparison of the pulse and lock-in thermography NDE techniques. NDT & E International, 41(7), 501-509.

[10] Balageas, D. L., & Roche, J. M. (2014). Common tools for quantitative time-resolved pulse and step-heating thermography–part I: theoretical basis. *Quantitative InfraRed Thermography Journal*, *11*(1), 43-56.

[11] Roche, J. M., & Balageas, D. L. (2015). Common tools for quantitative pulse and step-heating thermography–part II: experimental investigation. *Quantitative InfraRed Thermography Journal*, *12*(1), 1-23.

[12] Pickering, S. G., Chatterjee, K., Almond, D. P., & Tuli, S. (2013). LED optical excitation for the long pulse and lock-in thermographic techniques. *NDT & E International*, *58*, 72-77., Almond, D. P.,

[13] Almond, D. P., Angioni, S. L., & Pickering, S. G. (2017). Long pulse excitation thermographic non-destructive evaluation. *NDT & E International*, *87*, 7-14.
[14] Wang, Z., Tian, G., Meo, M., & Ciampa, F. (2018). Image processing-based quantitative damage evaluation in composites with long pulse thermography. *Ndt & E International*, *99*, 93-104.
[15] Kalyanavalli, V., Ramadhas, T. A., & Sastikumar, D. (2018). Long pulse thermography investigations of basalt fibre reinforced composite. *NDT & E International*, *100*, 84-91.
[16] Wang, Z., Zhu, J., Tian, G., & Ciampa, F. (2019). Comparative analysis of eddy current pulsed thermography and long pulse thermography for damage detection in metals and composites. *NDT & E International*, *107*, 102155.
[17] Mulaveesala, R., Vaddi, J. S., & Singh, P. (2008). Pulse compression approach to infrared nondestructive characterization. *Review of Scientific Instruments*, *79*(9).
Chatterjee, K., Roy, D., & tTuli, S. (2017). A novel pulse compression algorithm for frequency modulated active thermography using band-pass filter. *Infrared Physics & Technology*, *82*, 75-84.
[18] Gong, J., Liu, J., Qin, L., & Wang, Y. (2014). Investigation of carbon fiber reinforced polymer (CFRP) sheet with subsurface defects inspection using thermal-wave radar imaging (TWRI) based on the multi-transform technique. *Ndt & E International*, *62*, 130-136.
[19] Silipigni, G., Burrascano, P., Hutchins, D. A., Laureti, S., Petrucci, R., Senni, L., ... & Ricci, M. (2017). Optimization of the pulse-compression technique applied to the infrared thermography nondestructive evaluation. *NDT & E International*, *87*, 100-110.
[20] Laureti, S., Silipigni, G., Senni, L., Tomasello, R., Burrascano, P., & Ricci, M. (2018). Comparative study between linear and non-linear frequency-modulated pulse-compression thermography. *Applied optics*, *57*(18), D32-D39.
[21] Yang, R., He, Y., Mandelis, A., Wang, N., Wu, X., & Huang, S. (2018). Induction infrared thermography and thermal-wave-radar analysis for imaging inspection and diagnosis of blade composites. *IEEE Transactions on Industrial Informatics*, *14*(12), 5637-5647.
[22] Hedayatrasa, S., Poelman, G., Segers, J., Van Paepegem, W., & Kersemans, M. (2019). Novel discrete frequency-phase modulated excitation waveform for enhanced depth resolvability of thermal wave radar. *Mechanical Systems and Signal Processing*, *132*, 512-522.
[23] Luo, Z., Luo, H., Wang, S., Mao, F., Yin, G., & Zhang, H. (2022). The photothermal wave field and high-resolution photothermal pulse compression thermography for ceramic/metal composite solids. *Composite Structures*, *282*, 115069.
[24] Misaridis, T., & Jensen, J. A. (2005). Use of modulated excitation signals in medical ultrasound. Part I: Basic concepts and expected benefits. *IEEE transactions on ultrasonics, ferroelectrics, and frequency control*, *52*(2), 177-191.
[25] Hutchins, D., Burrascano, P., Davis, L., Laureti, S., & Ricci, M. (2014). Coded waveforms for optimised air-coupled ultrasonic nondestructive evaluation. *Ultrasonics*, *54*(7), 1745-1759.
[26] Burrascano, P., Laureti, S., Senni, L., & Ricci, M. (2018). Pulse compression in nondestructive testing applications: reduction of near sidelobes exploiting reactance transformation. *IEEE Transactions on Circuits and Systems I: Regular Papers*, *66*(5), 1886-1896.
[27] Turin, G. (1960). An introduction to matched filters. *IRE transactions on Information theory*, *6*(3), 311-329.
[28] Hedayatrasa, S., Poelman, G., Segers, J., Van Paepegem, W., & Kersemans, M. (2022). Phase inversion in (vibro-) thermal wave imaging of materials: Extracting the AC component and filtering nonlinearity. *Structural Control and Health Monitoring*, *29*(4), e2906.
[29] Budišin, S. Z. (1991). Efficient pulse compressor for Golay complementary sequences. *Electronics Letters*, *27*(3), 219-220.
[30] Garcia-Rodriguez, M., Yañez, Y., Garcia-Hernandez, M. J., Salazar, J., Turo, A., & Chavez, J. A. (2010). Application of Golay codes to improve the dynamic range in ultrasonic Lamb waves air-coupled systems. *NDT & e International*, *43*(8), 677-686.
[31] Arora, V., Mulaveesala, R., Kumar, S., & Wuriti, S. (2021). Non-destructive evaluation of carbon fiber reinforced polymer using Golay coded thermal wave imaging. *Infrared Physics & Technology*, *118*, 103908.
[32] Ricci, M., Senni, L., & Burrascano, P. (2012). Exploiting pseudorandom sequences to enhance noise immunity for air-coupled ultrasonic nondestructive testing. *IEEE Transactions on Instrumentation and Measurement*, *61*(11), 2905-2915.

[33] Malekmohammadi, H., Migali, A., Laureti, S., & Ricci, M. (2021). A pulsed eddy current testing sensor made of low-cost off-the-shelf components: Overview and application to pseudo-noise excitation. *IEEE Sensors Journal*, *21*(20), 23578-23587.
[34] Candoré, J. C., Bodnar, J. L., Detalle, V., & Grossel, P. (2012). Non-destructive testing of works of art by stimulated infrared thermography. *The European Physical Journal-Applied Physics*, *57*(2).
[35] Bodnar, J. L., Nicolas, J. L., Candoré, J. C., & Detalle, V. (2012). Non-destructive testing by infrared thermography under random excitation and ARMA analysis. *International Journal of Thermophysics*, *33*(10-11), 2011-2015.
[36] Vrabie, V., Perrin, E., Bodnar, J. L., Mouhoubi, K., & Detalle, V. (2012, August). Active ir thermography processing based on higher order statistics for nondestructive evaluation. In *2012 Proceedings of the 20th European Signal Processing Conference (EUSIPCO)* (pp. 894-898). IEE
[37] Dinan, E. H., & Jabbari, B. (1998). Spreading codes for direct sequence CDMA and wideband CDMA cellular networks. *IEEE communications magazine*, *36*(9), 48-54.
[38] Schroeder, M. R. (1979). Integrated-impulse method measuring sound decay without using impulses. *The Journal of the Acoustical Society of America*, *66*(2), 497-500.
[39] Sutter, E. E. (2001). Imaging visual function with the multifocal m-sequence technique. *Vision research*, *41*(10-11), 1241-1255.
[40] Golomb, S. W. (2017). *Shift register sequences: secure and limited-access code generators, efficiency code generators, prescribed property generators, mathematical models*. World Scientific.
[41] Sarwate, D. V., & Pursley, M. B. (1980). Crosscorrelation properties of pseudorandom and related sequences. *Proceedings of the IEEE*, *68*(5), 593-619.
[42] Hansen, T., & Mullen, G. L. (1992). Primitive polynomials over finite fields. *Mathematics of Computation*, *59*(200), 639-643.
[43] Golay, M. (1983). The merit factor of legendre sequences. *IEEE Transactions on Information Theory*, *29*(6), 934-936.
[44] Lu, R. T. M. A. C. (1989). *Algorithms for discrete Fourier transform and convolution*. Cham, Switzerland: Springer.
[45] Luke, H. D., Schotten, H. D., & Hadinejad-Mahram, H. (2003). Binary and quadriphase sequences with optimal autocorrelation properties: A survey. *IEEE Transactions on Information Theory*, *49*(12), 3271-3282.
[46] Lau, S. K., Almond, D. P., & Patel, P. M. (1991). Transient thermal wave techniques for the evaluation of surface coatings. *Journal of Physics D: Applied Physics*, *24*(3), 428.
[47] Oswald-Tranta, B. (2017). Time and frequency behaviour in TSR and PPT evaluation for flash thermography. *Quantitative InfraRed Thermography Journal*, *14*(2), 164-184.
[48] Sfarra, Stefano, et al. (2020). Low thermal conductivity materials and very low heat power: a demanding challenge in the detection of flaws in multi-layer wooden cultural heritage objects solved by pulse-compression thermography technique. *Applied Sciences* 10(12), 4233.